\documentclass[10pt]{iopart}

\usepackage{bm}
\usepackage{graphicx}
\expandafter\let\csname equation*\endcsname\relax

\expandafter\let\csname endequation*\endcsname\relax
\usepackage{mathtools} 
\usepackage{multirow}
\usepackage{epstopdf}
\usepackage{hyperref}
\usepackage[normalem]{ulem}
\hypersetup{
    colorlinks,%
    citecolor=blue,%
    linkcolor=blue,%
    urlcolor=blue
}
\usepackage{outlines}
\usepackage{tikz}\usetikzlibrary{petri}

\begin{document}

\title{ Charge and spin fluctuations in superconductors with intersublattice and interorbital interactions }
\author{Lauro B. Braz}
\address{
Instituto de F\'{\i}sica, Universidade de S\~ao Paulo, Rua do Mat\~ao 1371, S\~ao Paulo, S\~ao Paulo 05508-090, Brazil
}
\address{
 Department of Physics and Astronomy, Uppsala University, Box 516, S-751 20 Uppsala, Sweden
}
\author{George B. Martins}
\address{Instituto de F\'isica, Universidade Federal de Uberl\^andia, 
Uberl\^andia, Minas Gerais 38400-902, Brazil}
\author{Luis G.~G.~V. Dias da Silva}
\address{
Instituto de F\'{\i}sica, Universidade de S\~ao Paulo, Rua do Mat\~ao 1371, S\~ao Paulo, S\~ao Paulo 05508-090, Brazil
}

\begin{abstract}
    Multiband superconductors have featured one of the main challenges to achieve a comprehensive understanding of unconventional superconductivity. 
    Here, the multiband character is studied separately as orbital and sublattice degrees of freedom, as they have different effects for the superconducting and magnetic or charge orders. 
    We build on the framework of the matrix random-phase approximation (RPA), which accounts for the RPA Feynman diagrams and also vertex corrections, to treat the electron-electron interactions in an off-site degenerate Hubbard model. 
    As a result, systems without a sublattice degree of freedom tend to be dominated by spin fluctuations, while systems with multiple sublattice sites and orbitals have the charge fluctuations favored. 
    Finally, we explicitly demonstrate that the known suppression of the superconducting pairing strength $\lambda$ by spin fluctuations from repulsive interactions at zero momentum transfer $\boldsymbol{q}$ is countered by the finite-$\boldsymbol{q}$ pairing, which always improves $\lambda$. 
\end{abstract}


\submitto{\JPCM}

\maketitle

\ioptwocol



\section{Introduction}
\label{sec:introduction}
Superconductivity is a state of matter characterized by the coexistence of zero resistivity and the Meissner effect~\cite{Parks1969,Tinkham2004}. The fundamental fundamental building block of the superconducting state is a boson formed by two electrons composing a pair bound by an attraction mechanism (e.g. a Cooper pair \cite{Cooper1956}). 
In the so-called ``conventional'' superconductors (typically metals) described by BCS theory \cite{bardeenTheorySuperconductivity1957}, the pairing mechanism originates from an effective attractive interaction between electrons mediated by phonons in the lattice.
There are, however, several examples of materials which exhibit a superconducting phase whose pairing mechanism is \emph{not} of the type described by BCS theory. Examples of such ``unconventional'' superconductors include cuprate and iron pnictide materials, which can achieve critical temperatures ($T_c$) of up to 130 K \cite{schillingSuperconductivity130Hg1993}. The microscopic origin of unconventional superconductivity and the resulting ``high $T_c$'' behavior are among the main unsolved problems in condensed matter physics.


Repulsive electron-electron interactions were suggested as the driving mechanism for pairing in these unconventional superconductors \cite{scalapinoCommonThreadPairing2012a}.
For condensed matter purposes, electrons have three main properties: spin, charge, and momentum.
Therefore, interactions between electrons have both spin and charge parts, which cause \textit{fluctuations}, in the ground state of the system, that are associated with a momentum transfer vector $\boldsymbol{q}$ resulting from the difference between the momenta of two scattering Cooper pairs.
The earlier studies on the influence of spin fluctuations, causing $\boldsymbol{q}=0$ pair-scattering, showed that this channel was detrimental to the superconducting $T_c$ of $^3$He \cite{berkEffectFerromagneticSpin1966}.
On the other hand, taking into account both spin and charge fluctuations, at $\boldsymbol{q}\neq0$, the effects of paramagnons were later shown to favor spin-triplet pairing in $^3$He \cite{andersonAnisotropicSuperfluidityMathrm1973}. 
Paramagnons are short-lived and high-energy spin excitations occurring in materials that lack long-range magnetic order, 
presenting some form of \emph{local} magnetization instead. The realization of the importance 
of paramagnons was already a harbinger of the role played by the occurrence of a 
superconducting phase neighboring a magnetic phase in a series of future unconventional 
superconductors.
The success of the electron-electron interaction theory of charge- and spin-fluctuation paramagnons motivated its investigation from the perspective of the Hubbard model \cite{scalapinoWavePairingSpindensitywave1986,scalapinoFermisurfaceInstabilitiesSuperconducting1987a}. 
This model is expected to describe the metal-insulator transition, via doping, in the phase diagram of the cuprate superconductors, in which the spin fluctuations may account for the superconducting state (spin-singlet $d$-wave) and the neighboring antiferromagnetic (AFM) state. 

Subsequent works went beyond the Hubbard model, establishing spin and charge fluctuations as one of the main factors underlying the mechanism of unconventional superconductivity for the cuprates \cite{monienSpinExcitationsPairing1990,monienSpinChargeExcitations1990,monthouxWeakcouplingTheoryHightemperature1992,esirgenFluctuationexchangeTheoryGeneral1997,pinesSpinFluctuationModel1998,esirgenFluctuationExchangeAnalysis1998,takimotoStrongcouplingTheorySuperconductivity2004,monthouxSuperconductivityPhonons2007}.
Large off-site interactions were shown to favor charge instabilities in the Hubbard model \cite{esirgenMathitWavePairing1999} and in the extended Hubbard model \cite{roigRevisitingSuperconductivityExtended2022}.
Meanwhile, the discovery of organic superconductors ($T_c$ of up to 38 K \cite{takabayashiDisorderFreeNonBCSSuperconductor2009}) prompted research on charge and spin fluctuation models, including off-site interactions, which revealed possible spin-triplet states \cite{schmalianPairingDueSpin1998,kurokiSpintripletFwavelikePairing2001}.
Similarly, in the strontium ruthenate superconductors ($T_c$ of up to 1.5 K \cite{maenoSuperconductivityLayeredPerovskite1994}), electron-electron interactions were used to predict the spin-triplet state~\cite{takimotoOrbitalFluctuationinducedTriplet2000}, 
and, more recently, were shown to describe the knight shift observed in this family \cite{romerKnightShiftLeading2019,romerLeadingSuperconductingInstabilities2022}.
The sign-changing $s$-wave superconducting symmetry was proposed to appear on iron-based superconductors ($T_c$ of up to 55 K \cite{renSuperconductivityPhaseDiagram2008}), in  superconducting domes at the flanks of an AFM dome \cite{graserNeardegeneracySeveralPairing2009,kemperSensitivitySuperconductingState2010}.
Such prediction was experimentally observed later \cite{allanIdentifyingFingerprintAntiferromagnetic2015}.

More recently, spin and charge fluctuations were proposed to describe superconductivity in the nickelate family \cite{kreiselSuperconductingInstabilitiesStrongly2022}, found to be in the class of high-temperature superconductors with a $T_c$ of up to 80 K \cite{sunSignaturesSuperconductivity802023a}.
Spin and charge fluctuations from electron-electron interactions are also proposed to yield the unconventional superconductivity and charge order of vanadium kagom\'e materials \cite{wuHarmonicFingerprintUnconventional2020}.
These are multiband systems that require a careful track of the superconducting pairing symmetry, also called the superconducting pairing ``fingerprint" \cite{wuIdentificationSuperconductingPairing2019}.
Finally, inspired by the paramagnon theory of $^3$He, it was proposed that magic-angle twisted bilayer graphene (MATBG) has its unconventional superconductivity caused by attractive intra-sublattice effective electronic interactions \cite{wuIdentificationSuperconductingPairing2019,huangPseudospinParamagnonsSuperconducting2022}. 
The detailed structure of the interactions in MATBG were recently explored by the 
current authors, revealing a possible transition from magnetic to superconducting to 
magnetic state again, as the off-site interactions are tuned \cite{brazSuperconductivitySpinFluctuations2023}.

A common thread for most of the materials mentioned above is their multiband influence on the superconducting and charge or magnetic orders. We argue, however, that the multiband character should 
be decomposed into orbital and sublattice degrees of freedom.
The systems thought to have one relevant sublattice site are, for example, strontium ruthenate, and iron-based superconductors, while the cuprates and the nickelates have relevant oxygen orbitals in between the copper \cite{mattDirectObservationOrbital2018} or nickel atoms \cite{kreiselSuperconductingInstabilitiesStrongly2022}, making them three-sublattice systems.
Insterestingly, all these materials share proposals of spin insulating states \cite{kurokawaUnveilingPhaseDiagram2023,liElastocaloricDeterminationPhase2022,varignonCompletePhaseDiagram2017,martinelliPhaseDiagramsIronbased2016}.
Also, the vanadium kagom\'e and MATBG superconductors have three and two sublattice sites, respectively.
We show that the presence of more than one sublattice site is essential to the emergence of charge fluctuations and, consequently, charge orders, or spin-triplet superconductivity.
Indeed, some of these multisublattice systems host charge ordered states \cite{neupertChargeOrderSuperconductivity2022,jiangChargeOrderBroken2019,wiseImagingNanoscaleFermisurface2009}.
On the other hand, it is believed that the cuprates need two orbitals on the copper sites and one in each oxygen \cite{mattDirectObservationOrbital2018,watanabeUnifiedDescriptionCuprate2021}, the strontium ruthenates need at least three orbitals \cite{takimotoOrbitalFluctuationinducedTriplet2000}, and the iron pnictides need at least five orbitals \cite{graserNeardegeneracySeveralPairing2009} to describe their low-energy properties, while for the nickelates single-orbital \cite{kitataniNickelateSuperconductorsRenaissance2020} and multiorbital \cite{lechermannLateTransitionMetal2020} models are currently under debate \cite{karpManyBodyElectronicStructure2020}.
The vanadium kagom\'e superconductors need at least two orbitals to properly describe their non-interacting electronic properties \cite{wuHarmonicFingerprintUnconventional2020}, while the MATBG has a superlattice decoupling of its effective orbitals \cite{lopesdossantosGrapheneBilayerTwist2007}.
We conclude that a careful understanding of the influence of the orbital and sublattice degrees of freedom is essential to understand the phase diagrams of these systems, and to predict the phase diagrams of new correlated materials.

In this work, we build on the matrix random-phase approximation (RPA), which properly accounts for the RPA diagrams and also vertex corrections \cite{altmeyerRoleVertexCorrections2016}, to describe models with multiple orbital and/or multiple sublattice  degrees of freedom, including off-site Hubbard-like and exchange-like interactions, for any system with $N_s$ sublattice sites and $N_o$ orbitals.
We demonstrate that the number of matrix elements in the full four-body space, which in principle grows with the fourth power on the number of sublattice sites and orbitals $N_s^4N_o^4$, can be reduced to a quadratic growth $3N_{s}^{2}N_{o}^{2}-2N_{s}^{2}N_{o}$ by neglecting channels less important than the higher-order terms in perturbation theory and, therefore, cannot generate ordered states.

Our results indicate that systems without sublattice degree of freedom tend to be dominated by spin fluctuations, while the sublattice degree of freedom, when aligned with the multiorbital character, favors charge fluctuations.
Following the result by Berk and Schrieffer \cite{berkEffectFerromagneticSpin1966}, we show analytically that spin fluctuations are detrimental to the superconducting $T_c$ only if a zero momentum transfer peak dominates the pairing vertex, while finite momentum transfer only increases the pairing strength and, possibly, $T_c$ \footnote{See equation~\eqref{eq:lambda_lower_bound} and related discussion.}. 
The increase in the pairing strength can yield strong-coupling superconductivity. In addition, we show that the dominant
bare susceptibility matrix element should lack the combination of inversion symmetry and a zero momentum transfer peak for the pairing vertex to hold a leading spin-triplet superconducting instability.

This paper is organized as follows:
In Sec.~\ref{sec:model} we present the non-interacting model, the Hubbard-like and exchange-like interactions, which are renormalized by the matrix-RPA method in Sec.~\ref{sec:many-body}; in Sec.~\ref{sec:exact} we derive the analytical results based on linear algebra theorems and derive general conclusions on the influence of the number of orbitals and sublattice sites into the charge and spin fluctuations; in Sec~\ref{sec:strong_coupling}, we discuss the possible strong-coupling superconductivity scenarios in the many-body model.
We present our final conclusions in Sec.~\ref{sec:conclusions}.

\section{Matrix-RPA formalism for multiorbital and multisublattice systems }
\label{sec:model}
In this section, we construct the basis of our orbital- and sublattice-dependent matrix-RPA approach.
We consider noninteracting $H_0$ (quadratic) and two-body interacting $H_\text{int}$ (quartic) parts of the Hamiltonian $H=H_0+H_\text{int}$, which will be further coupled in the matrix-RPA pairing vertex calculations (Sec.~\ref{sec:many-body}).
This coupling adds many-body effects to the treatment of the system.

The noninteracting Hamiltonian in real space is given by
\begin{equation}
    H_0 = \sum_{ij}\sum_{\alpha\beta s p, \sigma} t^{\alpha s \beta p}_{ij} c^{\dagger}_{i\alpha s\sigma}c_{j\beta p\sigma},
    \label{eq:H0}
\end{equation}
where $c^{\dagger}_{i\alpha s \sigma}$ ($c_{i\alpha s \sigma}$) creates (annihilates) an electron with spin $\sigma$ in the sublattice site $\alpha$ of the $i$-th unit cell and in the orbital $s$.
$t^{\alpha s \beta p}$ is the tunneling constant connecting different sublattice sites and orbitals.
There are $s=1,2,3,\ldots,N_o$ orbitals and $\alpha=1,2,3,\ldots,N_s$ sublattice sites, totaling $N_b = N_oN_s$ spin-degenerate bands.

In our formulation of multiorbital and multisublattice interaction theory we consider Hubbard-like and exchange-like interactions.
The Hubbard-like terms are of onsite ($U_{0_{\alpha\alpha}}$) and long-range ($U_{m_{\alpha\beta}}$, where the $m_{\alpha\beta}$-th nearest-neighbor is such that $m_{\alpha\beta}>0$) type, such as the exchange-like onsite ($J_{0_{\alpha\alpha}}$) and long-range ($J_{m_{\alpha\beta}}$ with ${m_{\alpha\beta}}>0$) terms.
We start by considering a two-body Hamiltonian
\begin{equation}
    \begin{split}
    H_{\text{int}} & = 
    \sum_{ij}\sum_{\alpha\tilde{\alpha}\beta\tilde{\beta}}\sum_{sptd,\sigma_{g}}
    {\text{\c{c}}^{\alpha\beta}_{ij}}V_{\sigma_{1}\sigma_{2}\sigma_{3}\sigma_{4},ij}^{\alpha s\tilde{\alpha} p,\beta t\tilde{\beta} d}\delta_{\alpha\tilde{\alpha}}\delta_{\beta\tilde{\beta}}\\
    & \times c_{i\alpha s\sigma_{1}}^{\dagger}c_{j\beta t\sigma_{2}}^{\dagger}c_{j\tilde{\beta} d\sigma_{3}}c_{i\tilde{\alpha} p\sigma_{4}} \left(\delta_{\sigma_{1}\sigma_{3}}\delta_{\sigma_{2}\sigma_{4}}+\delta_{\sigma_{2}\sigma_{3}}\delta_{\sigma_{1}\sigma_{4}}\right)\; ,
    \end{split}
\label{eq:Hgeneral}
\end{equation}
where $s,p,t,d$ are orbital indices, $\sigma_g$ are spin indices, and $\alpha,\beta,\tilde{\alpha},\tilde{\beta}$ are sublattice indices.
The real space Coulomb integrals $V_{\sigma_{1}\sigma_{2}\sigma_{3}\sigma_{4},ij}^{\alpha s\tilde{\alpha} p,\beta t,\tilde{\beta} d}$ consider up to $w$ nearest-neighbors. The delta functions in spin indices ensure spin conservation.
The factor ${\text{\c{c}}}^{\alpha\beta}_{ij}$ is present to avoid double counting over the exchange of spin indices $s\sigma_1$ and $t\sigma_2$, and $d\sigma_3$ and $p\sigma_4$ when $\alpha=\beta$, and $i=j$, in which case it takes the value ${\text{\c{c}}}^{\alpha\alpha}_{ii}=1/4$, 
and ${\text{\c{c}}}^{\alpha\beta}_{ij}=1$ otherwise.

In momentum space, the Coulomb integrals depend on a momentum transfer $\boldsymbol{q}$ (omitted in the following notation, i.e., $V_{\sigma_{1}\sigma_{2}\sigma_{3}\sigma_{4}}^{\alpha s\tilde{\alpha} p,\beta t,\tilde{\beta} d}$) and they can be split into spin and charge parts \cite{graserNeardegeneracySeveralPairing2009}
\begin{equation}
    \begin{split}
    V_{\sigma_{1}\sigma_{2}\sigma_{3}\sigma_{4}}^{\alpha s\tilde{\alpha} p,\beta t\tilde{\beta} d} = 
    &-\frac{1}{2}\mathcal{V}^{\alpha s\tilde{\alpha} p}_{\beta t\tilde{\beta} d}\boldsymbol{\sigma}_{\sigma_{1}\sigma_{4}}\cdot\boldsymbol{\sigma}_{\sigma_{2}\sigma_{3}} \\
    & + \frac{1}{2} \mathcal{U}^{\alpha s\tilde{\alpha} p}_{\beta t\tilde{\beta} d}\delta_{\sigma_{2}\sigma_{3}}\delta_{\sigma_{1}\sigma_{4}},
    \end{split}
\label{eq:charg_spin_parts}
\end{equation}
where $\boldsymbol{\sigma}_{\sigma_{1}\sigma_{4}}$ denotes a Pauli vector. By identifying all the Hubbard-like and exchange-like terms in equation~\eqref{eq:Hgeneral}, one obtains the charge $\hat{\mathcal{U}}$ and spin $\hat{\mathcal{V}}$ interaction matrices.
These $N_b^2\times N_b^2$ matrices are given in terms of indices ${\alpha s\tilde{\alpha} p,\beta t\tilde{\beta} d}$ \cite{graserNeardegeneracySeveralPairing2009,wuHarmonicFingerprintUnconventional2020} and here we use a convenient basis by which these matrices turn out to be block diagonal.

Before proceeding, we define some useful quantities. 
If the terms $\alpha,\beta$ denote different sublattice sites, the definition of the scalar $U_{\alpha\alpha}=\sum_{m_{\alpha\alpha},l} U_{m_{\alpha\alpha}}e^{i\boldsymbol{\delta}_{m_{\alpha\alpha},l}\cdot\boldsymbol{q}}$ includes same-sublattice interaction elements, 
while $U_{\alpha\beta}=2\sum_{m_{\alpha\beta},l} U_{m_{\alpha\beta}}e^{i\boldsymbol{\delta}_{m_{\alpha\beta},l}\cdot\boldsymbol{q}}$ 
includes different-sublattice elements, where $\boldsymbol{\delta}_{m_{\alpha\alpha},l}$ ($\boldsymbol{\delta}_{m_{\alpha\beta},l}$) denotes the vector connecting same-sublattice (different-sublattice) $m_{\alpha\alpha}$-th ($m_{\alpha\beta}$-th) neighbors and $l$ is a $m_{\alpha\alpha}$($m_{\alpha\beta}$)-dependent quantity that runs over the sites at the same distance.
We make analogous definitions for $U_{\alpha\beta}',J_{\alpha\beta}$, and $J_{\alpha\beta}'$ and, conveniently, $M_{\alpha\beta}=-U'_{\alpha\beta}+2J_{\alpha\beta}$ and $N_{\alpha\beta}=2U'_{\alpha\beta}-J_{\alpha\beta}$.
For example, the usual relation $U'_{m_{\alpha\beta}}=U_{m_{\alpha\beta}}-2J_{m_{\alpha\beta}}$ \cite{graserNeardegeneracySeveralPairing2009} yields $M_{\alpha\beta}=-U_{\alpha\beta}+4J_{\alpha\beta}$ and $N_{\alpha\beta}=2U_{\alpha\beta}-3J_{\alpha\beta}$.
Since $U_{m_{\alpha\beta}}$ is typically larger than $J_{m_{\alpha\beta}}$, as the first depends on the square norm of the wave-functions, $N_{m_{\alpha\beta}},U'_{m_{\alpha\beta}}$ are positive definite, while $M_{m_{\alpha\beta}}$ can be either positive or negative, depending on material-specific properties. Reference~\cite{letouzeParametrizationCoulombInteraction2023} presents a thorough study of the properties of the Coulomb integrals in equation~\eqref{eq:Hgeneral}.

Thus, the charge interaction matrix takes the form (see  \ref{Sec:matrices})
\begin{equation}
    \hat{\mathcal{U}} = 
    \begin{bmatrix}
    \hat{\mathcal{U}}_{ss} & \hat{0} & \hat{0}\\
    \hat{0} & \hat{\mathcal{U}}_{sp} & \hat{0}\\
    \hat{0} & \hat{0} & \hat{0} 
    \end{bmatrix}
    , \hat{\mathcal{U}}_{ss} = 
    \begin{bmatrix}
    \hat{\mathcal{U}}_{AA}^{ss} & \hat{\mathcal{U}}_{AB}^{ss} & \hat{\mathcal{U}}_{AC}^{ss} & \cdots\\
    \hat{\mathcal{U}}_{AB}^{ss\dagger} & \hat{\mathcal{U}}_{BB}^{ss} & \hat{\mathcal{U}}_{BC}^{ss} & \cdots\\
    \hat{\mathcal{U}}_{AC}^{ss\dagger} & \hat{\mathcal{U}}_{BC}^{ss\dagger} &  \hat{\mathcal{U}}_{CC}^{ss} & \cdots\\
    \vdots & \vdots & \vdots & \ddots
    \end{bmatrix},
\end{equation}
\begin{equation}
    \hat{\mathcal{U}}_{sp} = \text{diag}(\hat{\mathcal{U}}^{sp},...,\hat{\mathcal{U}}^{sp}), 
    \hat{\mathcal{U}}^{sp} =
    \begin{bmatrix}
    \hat{\mathcal{U}}_{AA}^{sp} & \hat{\mathcal{U}}_{AB}^{sp} & \hat{\mathcal{U}}_{AC}^{sp} & \cdots\\
    \hat{\mathcal{U}}_{AB}^{sp\dagger} & \hat{\mathcal{U}}_{BB}^{sp} & \hat{\mathcal{U}}_{BC}^{sp} & \cdots\\
    \hat{\mathcal{U}}_{AC}^{sp\dagger} & \hat{\mathcal{U}}_{BC}^{sp\dagger} & \hat{\mathcal{U}}_{CC}^{sp} & \cdots\\
    \vdots & \vdots & \vdots & \ddots
    \end{bmatrix}, 
\end{equation}
\begin{equation}
    \hat{\mathcal{U}}_{\alpha\alpha}^{ss} = 
    \begin{bmatrix}
    U_{\alpha\alpha} & N_{\alpha\alpha} & N_{\alpha\alpha} & \cdots\\
    N_{\alpha\alpha} & U_{\alpha\alpha} & N_{\alpha\alpha} & \cdots\\
    N_{\alpha\alpha} & N_{\alpha\alpha} & U_{\alpha\alpha} & \cdots\\
    \vdots & \vdots & \vdots & \ddots
    \end{bmatrix},
    \hat{\mathcal{U}}_{\alpha\alpha}^{sp} = 
    \begin{bmatrix}
    J'_{\alpha\alpha} & M_{\alpha\alpha}\\
    M_{\alpha\alpha} & J'_{\alpha\alpha}
    \end{bmatrix},
\label{eq:Ussaa,spaa}
\end{equation}
\begin{equation}
    \hat{\mathcal{U}}_{\alpha\beta}^{ss} = 
    \begin{bmatrix}
    U_{\alpha\beta} & U'_{\alpha\beta} & U'_{\alpha\beta} & \cdots\\
    U'_{\alpha\beta} & U_{\alpha\beta} & U'_{\alpha\beta} & \cdots\\
    U'_{\alpha\beta} & U'_{\alpha\beta} & U_{\alpha\beta} & \cdots\\
    \vdots & \vdots & \vdots & \ddots
    \end{bmatrix},
    \hat{\mathcal{U}}_{\alpha\beta}^{sp} = 
    \begin{bmatrix}
    J'_{\alpha\beta} & J_{\alpha\beta}\\
    J_{\alpha\beta} & J'_{\alpha\beta}
    \end{bmatrix}.
\label{eq:Ussab,spab}
\end{equation}

In the above, $\hat{\mathcal{U}}$ has dimension $N_{b}^{2}\times N_{b}^{2}$, while $\hat{\mathcal{U}}_{ss}$ is an $N_{b}\times N_{b}$ matrix containing matrix elements in the $\alpha s\alpha s$ subspace, where the orbital index $s$ runs first for each sublattice $\alpha$.
We call $\hat{\mathcal{U}}_{ss}$ orbital density-density block.
$\hat{\mathcal{U}}_{\alpha\alpha}^{ss},\hat{\mathcal{U}}_{\alpha\beta}^{ss}$ are $N_{o}\times N_{o}$ matrices. 
$\hat{\mathcal{U}}_{sp}$ is an $N_{b}(N_{o}-1)\times N_{b}(N_{o}-1)$ containing $\alpha s\alpha p$ indices for $s\neq p$, where the indices $s,p$ run first for each $\alpha$ [the dimension is obtained noticing that there are $N_{o}(N_{o}-1)$ orbital permutations and $N_{s}$ sites].
In analogy to the density-density nomenclature \cite{graserNeardegeneracySeveralPairing2009}, refer to $\hat{\mathcal{U}}_{sp}$ as the ``orbital current-current'' block, because the interacting electrons are tunneling between orbitals ($s\neq p$) in these matrix elements.
$\hat{\mathcal{U}}^{sp}$ is a $2N_{s}\times2N_{s}$ matrix and there are $N_{o}(N_{o}-1)/2$ of those matrices (over two because each matrix includes two permutations of orbital indices). 
$\hat{\mathcal{U}}_{\alpha\alpha}^{sp},\hat{\mathcal{U}}_{\alpha\beta}^{sp}$ are $2\times2$ matrices composed by each $s,p$ exchange in the basis, \emph{i.e.}, $\alpha s\alpha p$ and $\alpha p\alpha s$. 

Analogously to the charge case, the spin interaction matrix is given by (see  \ref{Sec:matrices})
\begin{align}
    \hat{\mathcal{V}} & = 
    \begin{bmatrix}
    \hat{\mathcal{V}}_{ss} & \hat{0} & \hat{0} \\
    \hat{0} & \hat{\mathcal{V}}_{sp} & \hat{0} \\
    \hat{0} & \hat{0} & \hat{0}
    \end{bmatrix}
    , \hat{\mathcal{V}}_{ss} = 
    \begin{bmatrix}
    \hat{\mathcal{V}}_{AA}^{ss} & \hat{0} & \hat{0} & \cdots\\
    \hat{0} & \hat{\mathcal{V}}_{BB}^{ss} & \hat{0} & \cdots\\
    \hat{0} & \hat{0} & \hat{\mathcal{V}}_{CC}^{ss} & \cdots\\
    \vdots & \vdots & \vdots & \ddots
    \end{bmatrix},
\end{align}
\begin{align}
    \hat{\mathcal{V}}_{sp} & = \text{diag}(\hat{\mathcal{V}}^{sp},...,\hat{\mathcal{V}}^{sp}), 
    \hat{\mathcal{V}}^{sp} =
    \begin{bmatrix}
    \hat{\mathcal{V}}_{AA}^{sp} & \hat{0} & \hat{0} & \cdots\\
    \hat{0} & \hat{\mathcal{V}}_{BB}^{sp} & \hat{0} & \cdots\\
    \hat{0} & \hat{0} & \hat{\mathcal{V}}_{CC}^{sp} & \cdots\\
    \vdots & \vdots & \vdots & \ddots
    \end{bmatrix},
\end{align}
\begin{align}
    \hat{\mathcal{V}}_{\alpha\alpha}^{ss} & = 
    \begin{bmatrix}U_{\alpha\alpha} & J_{\alpha\alpha} & J_{\alpha\alpha} & \cdots\\
    J_{\alpha\alpha} & U_{\alpha\alpha} & J_{\alpha\alpha} & \cdots\\
    J_{\alpha\alpha} & J_{\alpha\alpha} & U_{\alpha\alpha} & \cdots\\
    \vdots & \vdots & \vdots & \ddots
    \end{bmatrix},
    \hat{\mathcal{V}}_{\alpha\alpha}^{sp} = 
    \begin{bmatrix}
    J'_{\alpha\alpha} & U'_{\alpha\alpha}\\
    U'_{\alpha\alpha} & J'_{\alpha\alpha}
    \end{bmatrix}.
\label{eq:Vssaa,spaa}
\end{align}

The full interaction space has dimension $N_{b}^{2}\times N_{b}^{2}$, which means $N_{b}^{4}$ momentum integrals performed to compute the bare susceptibility, as further discussed in Sec.~\ref{sec:many-body}.
This growth is too computationally expensive in terms of number of bands. 
However, we further show that the interaction matrices multiply the susceptibility in the pairing vertex, resulting in only two blocks with non-zero elements, one of dimension $N_{b}\times N_{b}$ ($\hat{\mathcal{U}}_{ss}$) and the other of dimension $N_{b}(N_{o}-1)\times N_{b}(N_{o}-1)$ ($\hat{\mathcal{U}}_{sp}$), where the latter has the structure of $N_{o}(N_{o}-1)/2$ non-zero blocks $\hat{\mathcal{U}}^{sp}$ of dimension $2N_{s}\times2N_{s}$. 
This reduces the pairing vertex non-zero elements from $N_{b}^{4}=N_{s}^{4}N_{o}^{4}$ to $N_{s}^{2}N_{o}^{2}+2N_{s}^{2}(N_{o}^{2}-N_{o})=3N_{s}^{2}N_{o}^{2}-2N_{s}^{2}N_{o}$. 
Under the approximation that the other zero terms on the interaction matrices cannot generate divergences on the susceptibilities, we conclude that the number of computations grows quadratically and not quartically as one would expect.

\section{ Coulomb interaction effects }
\label{sec:many-body}

In the following, we treat the Coulomb interaction effects, in the intermediate coupling regime, using the matrix-RPA formalism.
This is a temperature-dependent method for which the most divergent diagrams are taken into account, generating infinite cancellations and allowing semi-analytical expressions.
In this way, the method accounts for interactions in many-body systems by simple matrix product computations.
Indeed, the matrix-RPA approach was shown to hold diagrams beyond the usual RPA, also including vertex corrections \cite{altmeyerRoleVertexCorrections2016}.
We here extend the known formalism to include both orbital and sublattice degrees of freedom.

\subsection{ Charge and spin susceptibility }
\label{sec:chargespinsuscep}

The spin and charge susceptibilities are used to probe for the corresponding 
fluctuations in the system. 
They are calculated within matrix RPA by solving the RPA Dyson series expanded over powers of the interactions in equation~\eqref{eq:Hgeneral}.
The RPA approximation consists in considering only scattering events with the same scattering vector 
$\boldsymbol{q}$. The spin and charge susceptibilities are respectively given by \cite{takimotoStrongcouplingTheorySuperconductivity2004,graserNeardegeneracySeveralPairing2009,wuIdentificationSuperconductingPairing2019}
\begin{align}           
    \hat{\chi}_{\text{s}}(\boldsymbol{q},i\omega)	=\hat{\chi}(\boldsymbol{q},i\omega)\left[\hat{1}-\hat{\mathcal{V}}\hat{\chi}(\boldsymbol{q},i\omega)\right]^{-1}, \label{eq:chi_s} \\
    \hat{\chi}_{\text{c}}(\boldsymbol{q},i\omega) = \hat{\chi}(\boldsymbol{q},i\omega)\left[\hat{1}+\hat{\mathcal{U}}\hat{\chi}(\boldsymbol{q},i\omega)\right]^{-1}, \label{eq:chi_c}
\end{align} 
where the bare susceptibility matrix elements are given by
\begin{equation}
\begin{split}
    &\chi_{\beta t \tilde{\beta} d}^{\alpha s \tilde{\alpha} p}(\boldsymbol{q},i\omega)=\int_{0}^{\beta}\text{d}\tau\;e^{i\omega\tau}\sum_{\boldsymbol{k}\boldsymbol{k}'}\sum_{\sigma\tilde{\sigma}} \\
    &\langle T_{\tau}c_{\alpha s\sigma}^{\dagger}(\boldsymbol{k}+\boldsymbol{q},\tau)c_{\tilde{\alpha}p\sigma}(\boldsymbol{k},\tau)c_{\beta t\tilde{\sigma}}^{\dagger}(\boldsymbol{k}'-\boldsymbol{q},0)c_{\tilde{\beta}d\tilde{\sigma}}(\boldsymbol{k}',0)\rangle.
\end{split}
\label{eq:chi0_raw}
\end{equation}
The integral upper limit is $\beta=1/(k_BT)$, $i\omega$ are bosonic frequencies, $\tau$ is the imaginary time, and $T_\tau$ accounts for time-ordering.
In this way, the electron tunneling happens from $\tilde{\alpha} p \rightarrow \alpha s$ and $\tilde{\beta} d \rightarrow \beta t$.
This motivates the definition of the quantity $\chi(\boldsymbol{q},i\omega)=\sum_{\alpha s ,\beta p}\chi^{\alpha s \alpha s}_{\beta p \beta p}/2^4$ \cite{graserNeardegeneracySeveralPairing2009} as the bare homogeneous susceptibility, because these are the density-density elements in equation~\eqref{eq:chi0_raw} at $\boldsymbol{q}=0$.
Using Wick's theorem, equation~\eqref{eq:chi0_raw} can be shown to have the form of loop propagation (bubble) Feynman diagrams \cite{graserNeardegeneracySeveralPairing2009}
\begin{equation}
\begin{split}
    \chi_{\beta t \tilde{\beta} d}^{\alpha s \tilde{\alpha} p}(\boldsymbol{q},i\omega)=\\
    -\frac{1}{\beta N_{\boldsymbol{k}}}\sum_{\boldsymbol{k},i\omega_{n}}&G_{\tilde{\beta} d \alpha s}(\boldsymbol{k},i\omega_{n})G_{\tilde{\alpha} p\beta t}(\boldsymbol{k}+\boldsymbol{q},i\omega_{n}+i\omega),
\end{split}
\label{eq:chi0_GF}
\end{equation}
where $N_{\boldsymbol{k}}$ is the number of Brillouin zone (BZ) $\boldsymbol{k}$-points considered in the momentum summation and $i\omega_n=2\pi(n+1)/\beta$ are the fermionic Matsubara frequencies.
Thus, the propagations happen between states $\tilde{\beta} d\rightarrow\alpha s$ and $\tilde{\alpha}p\rightarrow\beta t$.
Counterintuitively, the momentum flow is different from the physical electron flow.

At this point, any non-interacting Green's function could be used in equation~\eqref{eq:chi0_GF}. 
For example, consider the one corresponding to the model of equation~\eqref{eq:H0}
\begin{equation}
    G_{\tilde{\beta} d \alpha s}(\boldsymbol{k},i\omega)=\sum_{\nu}\frac{a_{\nu}^{\tilde{\beta} d}(\boldsymbol{k})a_{\nu}^{\alpha s}(\boldsymbol{k})^{*}}{i\omega-E_{\nu}(\boldsymbol{k})},
\label{eq:GF}
\end{equation}
where $E_\nu$ are eigenvalues of $H_0$ and $a_{\nu}^{\beta t}$ their eigenvector coefficients.
Replacing equation~\eqref{eq:GF} into equation~\eqref{eq:chi0_GF}, summing over the fermionic Matsubara frequencies $i\omega_n$, and considering $i\omega\rightarrow\omega+i0^+$, one obtains \cite{graserNeardegeneracySeveralPairing2009,wuIdentificationSuperconductingPairing2019}
%
\begin{equation}
\begin{split}
\chi_{\beta t\tilde{\beta}d}^{\alpha s\tilde{\alpha}p}&(\boldsymbol{q},\omega)=\\
&-\frac{1}{N} \sum_{\boldsymbol{k},\nu\nu'}\frac{a_{\nu}^{\beta t}(\boldsymbol{k})a_{\nu}^{\alpha s}(\boldsymbol{k})^{*}a_{\nu'}^{\tilde{\alpha}p}(\boldsymbol{k}+\boldsymbol{q})a_{\nu'}^{\tilde{\beta}d}(\boldsymbol{k}+\boldsymbol{q})^{*}}{\omega+E_{\nu}(\boldsymbol{k})-E_{\nu'}(\boldsymbol{k}+\boldsymbol{q})+i0^{+}} \\
& \times \left[f\left(E_{\nu'}(\boldsymbol{k}+\boldsymbol{q})-(\omega+i0^{+})\right)-f\left(E_{\nu}(\boldsymbol{k})\right)\right],
\label{eq:chi0}
\end{split}
\end{equation}
which depends on the Fermi-Dirac distribution $f$ for a given temperature $T$.

The generalized Stoner criterion (namely, the vanishing of the denominator in Eqs.~\eqref{eq:chi_s} or \eqref{eq:chi_c} at $\omega=0$) establishes the condition for the transition between a paramagnetic (uniform density) state, possibly favoring a superconducting phase, and a magnetically (charge) ordered one, in the particle-hole channel, with a critical temperature higher than $T_c$ \cite{bickersConservingApproximationsStrongly1989b,bickersConservingApproximationsStrongly1991,takimotoStrongcouplingTheorySuperconductivity2004,sakakibaraOriginMaterialDependence2012}.
We define the spin ($\alpha_s$) and charge ($\alpha_c$) Stoner parameters by solving the following eigenvalue equations ~\cite{bickersConservingApproximationsStrongly1989b,bickersConservingApproximationsStrongly1991,takimotoStrongcouplingTheorySuperconductivity2004,engstromStraininducedSuperconductivityMathrm2023} 
\begin{equation}
\begin{split}
    & \hat{1}\alpha_s - \hat{\mathcal{V}}\hat{\chi} = 0, \\
    & \hat{1}\alpha_c + \hat{\mathcal{U}}\hat{\chi} = 0.
\end{split}
\label{eq:stoner}
\end{equation}
The Stoner criterion is achieved when $\alpha=\max \{ \alpha_s,\alpha_c \} =1$.
Hereafter, we will use the Stoner parameters as representatives of the relevance of spin and charge fluctuations \cite{sakakibaraOriginMaterialDependence2012}.

\subsection{ Multiorbital pairing vertex }
\label{sec:pairingvertex}

We proceed with the calculation of the multiorbital pairing vertex $\Gamma_{\beta t \tilde{\beta} d}^{\alpha s \tilde{\alpha} p,S}(\boldsymbol{k},\boldsymbol{k}',\omega)$ 
for scattering of a singlet pair $(\boldsymbol{k} \uparrow \alpha s, -\boldsymbol{k} \downarrow \beta t)$ 
with another pair $(\boldsymbol{k}' \uparrow \tilde{\alpha} p, -\boldsymbol{k}' \downarrow \tilde{\beta} d)$.
The spin-triplet pairing $\Gamma_{\beta t \tilde{\beta} d}^{\alpha s \tilde{\alpha} p,T}(\boldsymbol{k},\boldsymbol{k}',\omega)$ 
represents the scattering of a pair $(\boldsymbol{k} \sigma \alpha s, -\boldsymbol{k} \sigma \beta t)$ 
with another pair $(\boldsymbol{k}' \sigma \tilde{\alpha} p, -\boldsymbol{k}' \sigma \tilde{\beta} d)$.
The kernel function $\Gamma^{\eta}(\boldsymbol{k},\boldsymbol{k}')$ ($\eta=S,T$) is an $N_{\boldsymbol{k}}\times N_{\boldsymbol{k}'}$ matrix and it is related to the matrix-RPA charge and spin susceptibilities as \cite{scalapinoWavePairingSpindensitywave1986,esirgenFluctuationexchangeTheoryGeneral1997,schmalianPairingDueSpin1998,takimotoStrongcouplingTheorySuperconductivity2004,wuIdentificationSuperconductingPairing2019}
\begin{equation}
    \begin{split}
        \Gamma^{\eta}(\boldsymbol{k},\boldsymbol{k}') = &\sum_{stpq}\sum_{\alpha\tilde{\alpha}\beta\tilde{\beta}}a_{\nu_{-k}}^{ \beta t,*}(-\boldsymbol{k})a_{\nu_k}^{\alpha s,*}(\boldsymbol{k}) \\
        & \times \text{Re}[\Gamma_{\beta t \tilde{\beta} d}^{\alpha s \tilde{\alpha} p,\eta}(\boldsymbol{k},\boldsymbol{k}',0)]a_{\nu_{k'}}^{\tilde{\alpha} p}(\boldsymbol{k}')a_{\nu_{-k'}}^{\tilde{\beta} d}(-\boldsymbol{k}') \; ,\label{eq:kernel} \\
    \end{split}
\end{equation}
\begin{equation}
    \begin{split}
        \Gamma^{\alpha s \tilde{\alpha} p,\eta}_{\beta t \tilde{\beta} d}(\boldsymbol{k},\boldsymbol{k}',\omega) = \Bigg{[}\zeta_{\eta} & \hat{\mathcal{V}}\hat{\chi}_{s}(\boldsymbol{k}-\boldsymbol{k}',\omega)\hat{\mathcal{V}}+\frac{1}{2}\hat{\mathcal{V}} \\
        -\frac{1}{2} & \hat{\mathcal{U}}\hat{\chi}_{c}(\boldsymbol{k}-\boldsymbol{k}',\omega)\hat{\mathcal{U}}+\frac{1}{2}\hat{\mathcal{U}}\Bigg{]}_{\alpha s \beta t}^{\tilde{\beta} d \tilde{\alpha} p},\label{eq:vertex}
    \end{split}
\end{equation}
%
where $\zeta_{\eta=S}=3/2$ and $\zeta_{\eta=T}=-1/2$ denote the spin-singlet and spin-triplet channels, respectively, and $a_{\nu_{-k}}^{\tilde{\beta} d}$ are the $H_0$ eigenvector coefficients computed at the Fermi surface band $\nu$ corresponding to the momentum value $-\boldsymbol{k}$ \cite{graserNeardegeneracySeveralPairing2009,wuHarmonicFingerprintUnconventional2020}.
Since the pairing vertex in equation~\eqref{eq:vertex} always depends on the interaction matrices $\hat{\mathcal{U}}$ and $\hat{\mathcal{V}}$, we immediately notice that the pairing interaction only depends on the orbital density-density and orbital current-current indices, the non-zero blocks of the interaction matrices identified in Sec.~\ref{sec:model}.

The many-body interaction Hamiltonian we analyze results in the effective pairing interaction 

\begin{equation}
\begin{split}
    &V_{\text{eff}}^S(\omega) = \sum_{\boldsymbol{k}\boldsymbol{k}'}\sum_{sptd}\sum_{\alpha\tilde{\alpha}\beta\tilde{\beta}}\Gamma^{\alpha s \tilde{\alpha} p,S}_{\beta t \tilde{\beta} d}(\boldsymbol{k},\boldsymbol{k}',\omega) a_{\nu}^{\beta t,*}(-\boldsymbol{k}) \\
    &\times a_{\nu}^{\alpha s,*}(\boldsymbol{k}) a_{\nu}^{\tilde{\alpha} p}(\boldsymbol{k}')a_{\nu}^{\tilde{\beta} d}(-\boldsymbol{k}') c_{\boldsymbol{k}\alpha s\uparrow}^{\dagger}c_{-\boldsymbol{k}\beta t\downarrow}^{\dagger}c_{-\boldsymbol{k}'\tilde{\beta }d\downarrow}c_{\boldsymbol{k}'\tilde{\alpha} p\uparrow} 
\end{split}
\label{eq:Veff_S}
\end{equation}
for the spin-singlet channel and
\begin{equation}
\begin{split}
    &V_{\text{eff}}^T(\omega) = \sum_{\boldsymbol{k}\boldsymbol{k}'}\sum_{sptd,\sigma}\sum_{\alpha\tilde{\alpha}\beta\tilde{\beta}} \Gamma^{\alpha s \tilde{\alpha} p,T}_{\beta t \tilde{\beta} d}(\boldsymbol{k},\boldsymbol{k}',\omega) a_{\nu}^{\beta t,*}(-\boldsymbol{k}) \\
    &\times a_{\nu}^{\alpha s,*}(\boldsymbol{k}) a_{\nu}^{\tilde{\alpha} p}(\boldsymbol{k}')a_{\nu}^{\tilde{\beta} d}(-\boldsymbol{k}') c_{\boldsymbol{k}\alpha s\sigma}^{\dagger}c_{-\boldsymbol{k}\beta t\sigma}^{\dagger}c_{-\boldsymbol{k}'\tilde{\beta }d\sigma}c_{\boldsymbol{k}'\tilde{\alpha} p\sigma} 
\end{split}
\label{eq:Veff_T}
\end{equation}
for the spin-triplet channel. 
These particle-particle channel interacting Hamiltonians are more general than equation~\eqref{eq:Hgeneral} as they combine all the intersublattice interactions, as well as Hubbard and exchange terms, and the non-interacting electronic structure, in one many-body interaction. 
Analogous Hamiltonians can be derived for the particle-hole channel.

Finally, the pairing strength $\lambda_{\boldsymbol{k}}^{\eta}$ and the associated gap symmetry functional $\boldsymbol{g}^\eta(\boldsymbol{k}')$ are computed by solving the eigenvalue equation \cite{scalapinoWavePairingSpindensitywave1986,graserNeardegeneracySeveralPairing2009,wuIdentificationSuperconductingPairing2019}
\begin{equation}
    -\sum_{\boldsymbol{k}'}\oint_{C_{\boldsymbol{k}'}}\frac{d\boldsymbol{k}'_{||}}{v_{F}(\boldsymbol{k}')}\frac{1}{(2\pi)^{2}} \tilde{\Gamma}^{\eta}(\boldsymbol{k},\boldsymbol{k}')\boldsymbol{g}^\eta(\boldsymbol{k}') = \lambda_{\boldsymbol{k}}^{\eta} \boldsymbol{g}^\eta(\boldsymbol{k}),
\label{eq:pairing_strength}
\end{equation} 

%
in which $2\tilde{\Gamma}^{S}(\boldsymbol{k},\boldsymbol{k}') = \Gamma^{S}(\boldsymbol{k},\boldsymbol{k}')+\Gamma^{S}(\boldsymbol{k},-\boldsymbol{k}')$ [$2\tilde{\Gamma}^{T}(\boldsymbol{k},\boldsymbol{k}') = \Gamma^{T}(\boldsymbol{k},\boldsymbol{k}') - \Gamma^{T}(\boldsymbol{k},-\boldsymbol{k}')$] is the symmetric (antisymmetric) part of the kernel function $\Gamma^{S}(\boldsymbol{k},\boldsymbol{k}')$ [$\Gamma^{T}(\boldsymbol{k},\boldsymbol{k}')$] \cite{schmalianPairingDueSpin1998}.
We define the pairing strength $\lambda_{\eta} = \max \lambda_{\boldsymbol{k}}^{\eta}$, which relates to the higher critical temperature associated with the gap symmetry $\boldsymbol{g}^\eta(\boldsymbol{k})$.
In a BCS-like approach, for example, $T_c\propto e^{-1/\lambda}$, where $\lambda = \max\{ \lambda_S, \lambda_T \}$ \cite{scalapinoFermisurfaceInstabilitiesSuperconducting1987a}. 

The frequency-independent term $\hat{\mathcal{V}}/2+\hat{\mathcal{U}}/2$ in equation~\eqref{eq:vertex} is the mean-field, Hartree-Fock term, while the susceptibilities are a consequence of higher-order corrections to the particle-particle pairing vertex \cite{zhangTheorySpinfluctuationInduced2011}.
At the same time, these susceptibilities, when divergent, are representatives of the particle-hole orders \cite{linTwodimensionalHubbardModel1987,boehnkeSusceptibilitiesMaterialsMultiple2015}.
The fact that the particle-particle pairing vertex depends on a particle-hole-related quantity establishes an intricate relationship between superconductivity and particle-hole (spin and charge) fluctuations, giving origin to several possible phase diagrams.

\begin{figure}[t]
\begin{center}
\includegraphics[width=0.95\linewidth]{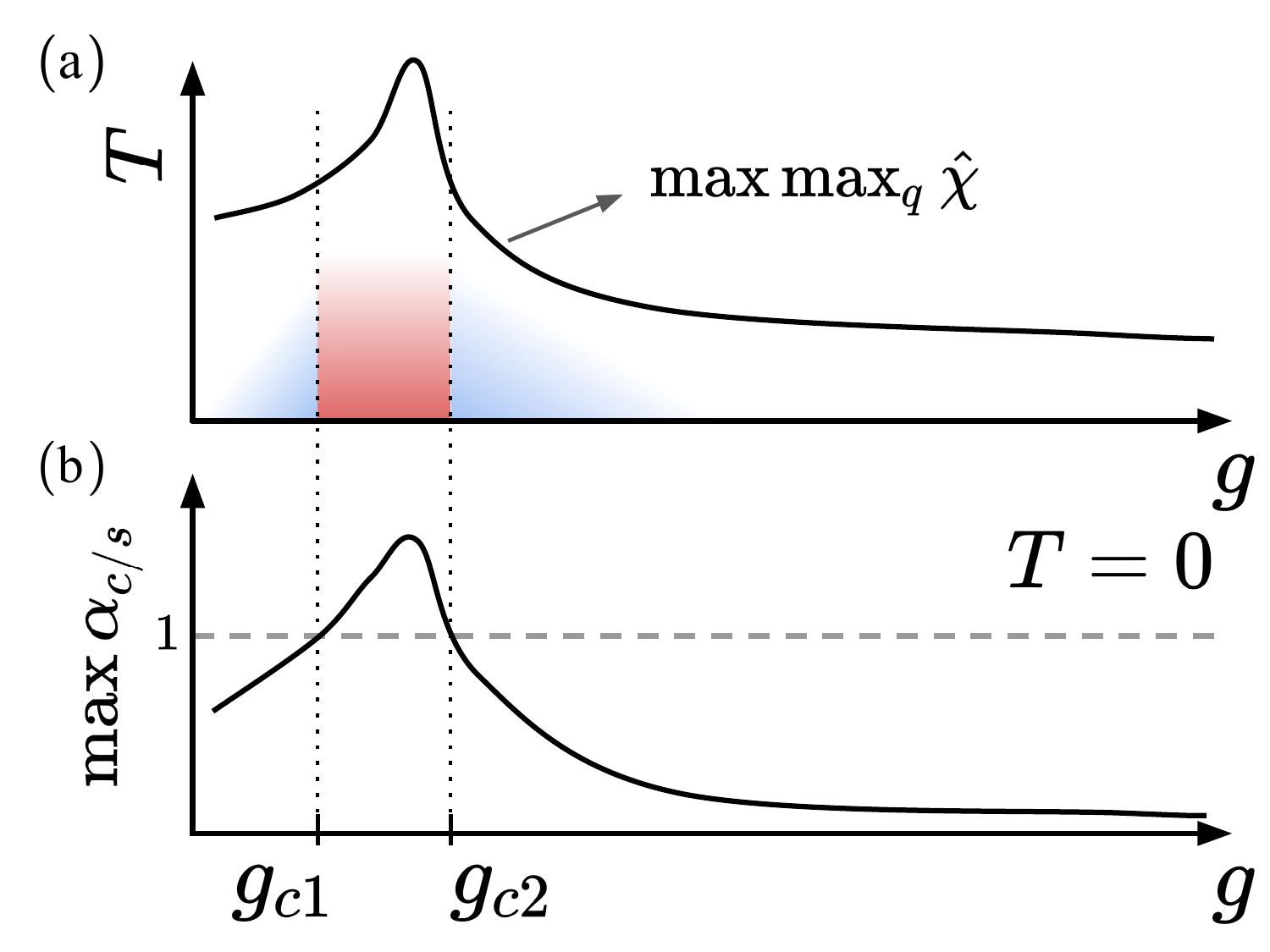}
\end{center}
\caption{ 
    (a) Qualitative phase diagram showing a charge or spin ordered state (red region), which is located between critical control parameter values, $g_{c1}$ and $g_{c2}$, and superconductivity (blue regions).
    Also, we show the $T=0$ qualitative curve of the main 
    bare susceptibility matrix element at the nesting vector leading the fluctuations ($\max\max_{\boldsymbol{q}}\hat{\chi}$) at each $g$. 
    (b) Main Stoner parameter of the respective phase at $T=0$ as a function of the control parameter $g$ for constant interaction parameters. $g_{c1}$ and $g_{c2}$ depend on the many-body interactions. A spin or charge order is present whenever $\max\alpha_{s/c}\geq1$. 
}
\label{fig:dome}
\end{figure}

Charge or spin fluctuations induce an ordered state in the particle-hole channel when the Stoner parameter in equation~\eqref{eq:stoner} $\alpha = \max\{\alpha_c,\alpha_s\}\geq 1$.
The order is associated with a divergence on the pairing vertex and has a critical temperature that is higher than the 
superconducting $T_c$ \cite{takimotoStrongcouplingTheorySuperconductivity2004,sakakibaraOriginMaterialDependence2012,bickersConservingApproximationsStrongly1989b,bickersConservingApproximationsStrongly1991}.
The fluctuation scenario is sensitive to the electronic structure of the system through the scattering momentum dependence of the bare susceptibility.
Consider, for example, the profile shown in figure~\ref{fig:dome}(a), in which the bare susceptibility varies as a function of a given control parameter $g$, e.g., pressure, doping, electric gating, displacement field, magnetic field, etc. 
There may be regions $g_{c1}<g<g_{c2}$ where the Stoner criterion is fulfilled [figure~\ref{fig:dome}(b)].
For $g<g_{c1}$ or $g>g_{c2}$, only fluctuations are present and those might cause a finite pairing vertex to present a nonzero pairing strength $\lambda$ and, therefore, nonzero $T_c$.
In this perspective, $g_{c1}$ and $g_{c2}$ are $T=0$ transitions. 
A second-order phase transition separates the superconducting, charge-, or spin-ordered states from the normal state, while a first-order phase transition connects the (possible) superconducting region that overlaps with the charge- or spin-ordered region \cite{lothmanUniversalPhaseDiagrams2017c}.


\section{Exact results for the Stoner parameters}
\label{sec:exact}

In order to express the possible fluctuation scenarios resulting from the spin-charge competition, we consider some simplifications and then derive bounds for the maximum spin and charge Stoner parameters.
If a matrix has real eigenvalues, the maximum eigenvalue is bounded by the deviation, given by the variance, from the trace of the matrix averaged over the space dimension (see Theorem 2.1 in 
Ref.~\cite{wolkowiczBoundsEigenvaluesUsing1980}).
The main assumption of this formal result is that Eqs.~\eqref{eq:stoner} must have only real eigenvalues, which is always true.
Aiming to obtain comprehensive analytical results, we define the susceptibility under the diagonal approximation \cite{kemperSensitivitySuperconductingState2010} in the same basis as the spin and charge matrices.
We further simplify the bare susceptibility by considering all orbital density-density terms and, in parallel, all orbital current-current terms as being the same, taking some characteristic value representing the most diverging bare susceptibility matrix element in each one of these two matrix blocks.
We show the full derivation of the bounds in  \ref{Sec:trace_bounds} and summarize the results in Table~\ref{tab:fluctuation_bounds}.

\begin{table*}[ht!]
\centering
\caption{ Analytical results for the bounds of the charge and spin Stoner parameters $\alpha_c$ and $\alpha_s$, respectively. Here, we show the conditions for which either the spin or charge fluctuations completely dominate the fluctuation scenario. The scattering momentum is taken as the nesting vector. }
\begin{tabular}{c|c|c}
                                                 & \begin{tabular}[c]{@{}c@{}}Leading\\ channel\end{tabular} & Condition                                                                                                                                                                                                            \\ \hline
\multirow{4}{*}{$\max \alpha_c > \max \alpha_s$} & \multirow{2}{*}{$\chi^{ss}_{\alpha\alpha}$}                             & $(2U'_{\alpha\alpha}-J_{\alpha\alpha})^{2}\frac{(N_{o}-1)}{(N_{b}-1)}+\frac{1}{(N_{b}-1)}\sum_{\alpha>\beta}(|U_{\alpha\beta}|^{2}+|U'_{\alpha\beta}|^{2}(N_o-1))>$ \\ 
        & & $(2U_{\alpha\alpha}+J_{\alpha\alpha}(N_{o}-1))^{2}$ \\ \cline{2-3}
                                                 & \multirow{2}{*}{$\chi^{sp}_{\alpha\alpha}$}                               & ($U'_{\alpha\alpha}+2J_{\alpha\alpha})^{2}+\sum_{\alpha>\beta}(|J'_{\alpha\beta}|^{2}+|J_{\alpha\beta}|^{2})>$ \\ 
                                                 & & $(2N_{s}-1)(2J'_{\alpha\alpha}+U'_{\alpha\alpha})^{2}$                                                    \\ \hline
\multirow{4}{*}{$\max\alpha_s>\max\alpha_c$}     & \multirow{2}{*}{$\chi^{ss}_{\alpha\alpha}$}                                & $(2U_{\alpha\alpha}+J_{\alpha\alpha})^{2}>(2U'_{\alpha\alpha}-J_{\alpha\alpha})^{2}(N_{b}-1)(N_{o}-1)$ \\
                                                & & $+(N_{b}-1)\sum_{\alpha>\beta}(|U_{\alpha\beta}|^{2}+|U'_{\alpha\beta}|^{2}(N_o-1))$                             \\ \cline{2-3}
                                                 & \multirow{2}{*}{$\chi^{sp}_{\alpha\alpha}$}                                & $(2J'_{\alpha\alpha}+U'_{\alpha\alpha})^{2}>$ \\
                                                 & & $(2N_{s}-1)(-U'_{\alpha\alpha}+2J_{\alpha\alpha})^{2}+(2N_{s}-1)\sum_{\alpha>\beta}(|J'_{\alpha\beta}|^{2}+|J_{\alpha\beta}|^{2})$  \\ \hline                                     
\end{tabular}
\label{tab:fluctuation_bounds}
\end{table*}
%

Table~\ref{tab:fluctuation_bounds} shows the conditions for the charge (spin) channel to lead over the spin (charge) channel, $\max \alpha_{c(s)}>\max\alpha_{s(c)}$, when the same-sublattice orbital density-density $\chi^{ss}_{\alpha\alpha}$ or the same-sublattice orbital current-current $\chi^{sp}_{\alpha\alpha}$ bare susceptibility channel dominates the fluctuation scenario, i.e., it diverges first when interactions are 
turned on. 
Notice that the maximum value has an explicit dependence on the number of orbitals $N_o$ and sublattice sites $N_s$ as well as bands $N_b=N_oN_s$.
In general, the interplay scenario, where the bounds for spin and charge Stoner parameters intersect, is favored when the number of orbitals and sublattice sites are large.
For example, when $N_o=N_s=1$, the spin fluctuations always dominate over the charge ones.
As the number of orbitals increases, but only one sublattice site is kept, the dimension of the matrices increases, thus the bounds get wider, losing precision, but still the upper bound for the spin fluctuations remains higher than the charge one (see \ref{Sec:trace_bounds}).
This suggests that spin fluctuations tend to dominate over charge fluctuations in systems without a sublattice degree of freedom.
Examples of this trend are the cuprates \cite{monthouxTheoryHightemperatureSuperconductivity1991,sakakibaraTwoOrbitalModelExplains2010,sakakibaraOriginMaterialDependence2012} and iron pnictides \cite{graserNeardegeneracySeveralPairing2009}.
The first one is believed to be a three-orbital system, while the second, a five-orbital one, but in both the dominant phase, 
close to the superconducting one, is antiferromagnetic.

%
\begin{table}[ht]
\centering
\caption{ Number of neighbors of an $A$ sublattice site for different lattice geometries.
$AA$ denotes same-sublattice and $AB,AC$ the different-sublattice neighbors.
The distance $D_{w\!=\!5}$ to the farthest unit cell within the $w=5$ nearest neighbor for each of these lattice geometries depends on the number of sublattice sites.}
\begin{tabular}{c|cccc}
                                                              & AA & AB & AC & \begin{tabular}[c]{@{}c@{}} $D_{w\!=\!5}$ \\ \end{tabular} \\ \hline
square                                                        & 28 &    &    & 3                                                                                                        \\
triangular                                                    & 36 &    &    & 3                                                                                                        \\
hexagonal                                                     & 12 & 12 &    & 2                                                                                                        \\
\begin{tabular}[c]{@{}c@{}}3-sublattice\\ square\end{tabular} & 8  & 8  & 8  & 1                                                                                                        \\
kagomé                                                        & 6  & 10 & 10 & 1                                                                                                       
\end{tabular}
\label{tab:lattice}
\end{table}
%

On the other hand, as the number of sublattice sites and orbitals increases, the charge fluctuations become more important than the spin fluctuations, depending on the lattice geometry.
As seen in Table~\ref{tab:fluctuation_bounds}, to have the condition $\max\alpha_c>\max\alpha_s$ fulfilled, one must add more sublattice sites to make the sublattice summation larger.
However, $U_{\alpha\alpha}$ are typically larger than $U_{\alpha\beta}$, because of the onsite term inside the first, such that one also needs a multiorbital system to have nonzero $U'_{\alpha\beta}$, guaranteeing charge dominance.
In their turn, the terms $U_{\alpha\beta}, U_{\alpha\beta}',J_{\alpha\beta},J_{\alpha\beta}'$ grow with the number of neighbors at each constant distance $\boldsymbol{r}_{i\alpha,j\beta}$ between sublattice site $\alpha$ at lattice site $i$ and sublattice site $\beta$ at lattice site $j$, where $\alpha\neq\beta$.
For example, up to $w=5$ nearest-neighbors, the number of different-sublattice neighbors of the hexagonal lattice is equal to the same-sublattice number of neighbors, as shown in table~\ref{tab:lattice}.
The three-sublattice square lattice (for example, when also 
considering oxygen in the cuprates) shows more different-sublattice sites than same-sublattice sites, as it is also the case in the kagom\'e 
lattice.

\begin{figure}[t]
\begin{center}
\includegraphics[width=0.95\linewidth]{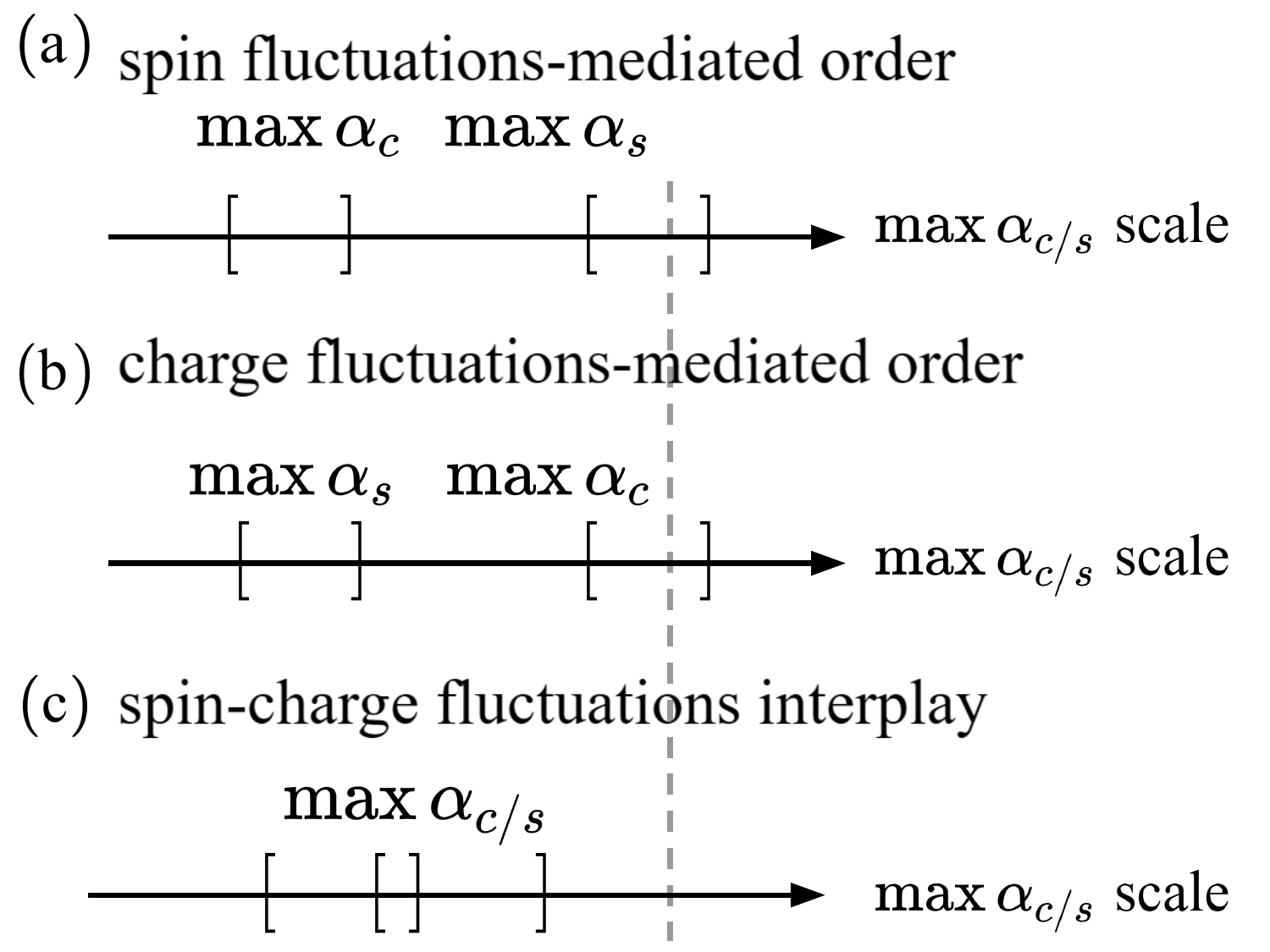}
\end{center}
\caption{ 
    Possible scenarios of interplay between charge and spin fluctuations. Panels (a)-(c) show the interval of possible main charge or spin Stoner parameters. The cases derived analytically consist in panels (a) and (b), while their simultaneous breaking is represented in panel (c).
    The vertical grey dashed line marks the divergence of the pairing vertex and, consequently, the entrance in an ordered state in the particle-hole channel.
}
\label{fig:fluctuations}
\end{figure}

In summary, there are three distinct situations in terms of the interplay between spin and charge fluctuations, which we depict in figure~\ref{fig:fluctuations}.
The situation shown in figure~\ref{fig:fluctuations}(a), where the spin fluctuations dominate over the charge fluctuations, is favored when either 
the number of sublattice sites or orbitals is low (notice that more sublattice sites $N_s$ and orbitals $N_o$ increase terms at right-hand side of the $\max\alpha_s>\max\alpha_c$ inequalities in table~\ref{tab:fluctuation_bounds}).
Figure~\ref{fig:fluctuations}(b) presents the scenario where the charge fluctuations dominate, which might happen when there are enough sublattice sites and more than one orbital (notice that the left-hand side of the $\max\alpha_c>\max\alpha_s$ inequalities is favored by intersublattice $|U_{\alpha\beta}|$ and intersublattice+interorbital $|U_{\alpha\beta}'|$ interactions). 
However, a lattice geometry-dependent limit cannot be exceeded, because the number of orbitals favors the spin fluctuations the most ($\max\alpha_s>\max\alpha_c$ inequalities) as the upper limit of $\max \alpha_s$ grows faster with the number of orbitals.
The need for off-site interactions to generate charge order has been discussed for the two-sublattice sites $\theta$-type organic superconductors \cite{tanakaSuperconductivityDueCharge2004,kimuraSpintripletSuperconductivityInduced2008}, in contrast to the onsite result, where spin fluctuations dominate \cite{schmalianPairingDueSpin1998}.
That can be attributed to the fact these calculations consider higher number of neighbors, in the triangular lattice, in comparison to the square lattice (table~\ref{tab:lattice}), increasing the $U_{\alpha\beta}$ on the bounds shown in table~\ref{tab:fluctuation_bounds}.
However, this phenomenon is more pronounced when considering more than one orbital.
Finally, the increasing number of both sublattice sites and 
orbitals can favor a scenario where the charge fluctuation bounds 
become wider, while the increasing number of orbitals broadens 
both spin and charge fluctuation bounds. 
In this case, the charge and spin fluctuations can compete 
and may cause both a suppression or an excessive increase of fluctuations, such that an ordered spin or charge ground-state is achieved [figure~\ref{fig:fluctuations}(c)]. 

The consequences of the spin-charge transition are essential for the underlying pairing mechanism driving superconductivity in the system. Indeed, 
as seen in equation~\eqref{eq:vertex}, the spin-singlet pairing vertex has a repulsive $\zeta_S=3/2$ factor on the spin susceptibility, while the spin-triplet solution has an attractive factor $\zeta_T=-1/2$.
On the other hand, the charge part has the same attractive factor $-1/2$ for both singlet and triplet cases. 
Repulsive singlet states ($\zeta_T=3/2$) are thus favored by dominating spin fluctuations, in contrast to charge fluctuations that favor the attractive triplet states over the singlet states because of the negative sign of both spin and charge fluctuations ($\zeta_T=-1/2$).
Therefore, multiorbital and multisublattice systems that favor charge fluctuations over spin fluctuations are more likely to be spin-triplet 
superconductors.
Thus, transitions from singlet to triplet might be possible under tuning of interactions, changing the bounds in Table~\ref{tab:fluctuation_bounds}.
Another possibility is to have a competition between spin and charge fluctuations [figure~\ref{fig:fluctuations}(c)], possibly causing a competition between spin-singlet and spin-triplet superconducting symmetries. 
As previously discussed, this situation can suppress both spin and charge fluctuations, lowering the pairing strength and, consequently, $T_c$. 
This means that the spin singlet and triplet competition is possibly detrimental to superconductivity \cite{aaseConstrainedWeakcouplingSuperconductivity2023a}. 

\section{Strong-coupling superconductivity from spin and charge fluctuations}
\label{sec:strong_coupling}

We explore here the possible origin of strong-coupling superconductivity driven by spin and charge fluctuations in multiorbital and multisublattice systems: unconventional superconductivity from finite nesting vectors.
In the context of equation~\eqref{eq:pairing_strength}, larger pairing strength $\lambda$ means potentially higher superconducting critical temperatures.
These originate from two distinct types of scattering $\boldsymbol{q}=\boldsymbol{k}-\boldsymbol{k}'$ events, viz., $\boldsymbol{q}=0$ ($\boldsymbol{k}=\boldsymbol{k}'$) and $\boldsymbol{q}\neq 0$ ($\boldsymbol{k}\neq\boldsymbol{k}'$).
Depending on the electronic structure, these events may generate nesting vectors for the magnetic- or charge-ordered phase, if realized.
We demonstrate in \ref{Sec:trace_bounds} that the lower bound of the pairing strength is given by 
\begin{equation}
    \begin{split}
        \lambda\geq-\sum_{\boldsymbol{k}}\frac{d\boldsymbol{k}_{||}}{v_{F}(\boldsymbol{k})}\frac{1}{N_{\boldsymbol{k}}(2\pi)^{2}}\Gamma(\boldsymbol{k},\boldsymbol{k})&\\
        +\Bigg[\frac{1}{N_{\boldsymbol{k}}(N_{\boldsymbol{k}}-1)}\sum_{\boldsymbol{k},\boldsymbol{k}'}\frac{d\boldsymbol{k}'_{||}d\boldsymbol{k}_{||}}{v_{F}(\boldsymbol{k}')v_{F}(\boldsymbol{k})}\frac{1}{(2\pi)^{4}}\Gamma(\boldsymbol{k},\boldsymbol{k}')\Gamma(\boldsymbol{k}',\boldsymbol{k})&\\
        -\frac{1}{N_{\boldsymbol{k}}^{2}(N_{\boldsymbol{k}}-1)}\left(\sum_{\boldsymbol{k}}\frac{d\boldsymbol{k}{}_{||}}{v_{F}(\boldsymbol{k})}\frac{1}{(2\pi)^{2}}\Gamma(\boldsymbol{k},\boldsymbol{k})\right)^{2}\Bigg]^{1/2}.&
    \end{split}
    \label{eq:lambda_lower_bound}
\end{equation}
%

The first term on the right-hand side of equation \eqref{eq:lambda_lower_bound} is a same-momentum term, i.e. $\boldsymbol{q}=0$, and it is accompanied by a minus sign.
Thus, $\boldsymbol{q}=0$ spin fluctuations decrease the pairing strength upon repulsive interactions \cite{berkEffectFerromagneticSpin1966}.
On the other hand, the finite-momentum ($\boldsymbol{k}\neq\boldsymbol{k}'$) pairing term in equation \eqref{eq:lambda_lower_bound} is positive, thus, any finite (even \textit{repulsive}) pairing vertex component with $\boldsymbol{q}\neq 0$ increases the pairing strength $\lambda$.
Now, as the system approaches the $g_{c1}$ or $g_{c2}$ transition points (see Fig.~\ref{fig:dome}), the denominator in the susceptibilities [equations~\eqref{eq:chi_s} and \eqref{eq:chi_c}] tend to vanish, and, 
as a consequence, the pairing vertex [equation \eqref{eq:vertex}], 
tend to diverge, causing an increase on the pairing strength 
$\lambda$. Therefore, finite momenta scattering 
events are a key aspect of strong coupling superconductivity from spin and 
charge fluctuations.
This aspect has been tangentially studied in the $(\pi,\pi)$ nesting of iron-based superconductors \cite{graserNeardegeneracySeveralPairing2009,luoNeutronARPESConstraints2010,martinsRPAAnalysisTwoorbital2013a,nicholsonCompetingPairingSymmetries2011,nicholsonRoleDegeneracyHybridization2011}, 
in the several nesting vectors of kagom\'e superconductors \cite{wuNatureUnconventionalPairing2021}, and in magic-angle twisted bilayer graphene \cite{wuHarmonicFingerprintUnconventional2020,brazSuperconductivitySpinFluctuations2023}. 



\section{Conclusions}
\label{sec:conclusions}
We have shown that an appropriate accounting of orbital and sublattice degrees of freedom leads to a comprehensive understanding of the influence of electron-electron interactions for superconducting, charge, and magnetic orders. 
Towards that end, we derived bounds for different scenarios of spin and charge fluctuations by using a degenerate Hubbard model as the interaction matrix in a perturbative expansion under the matrix-RPA, which accounts for the RPA and vertex corrections \cite{altmeyerRoleVertexCorrections2016}.
These bounds showed that increasing the number of sublattice sites relevant to the electronic structure favors charge fluctuations, mainly in multiorbital systems.
On the other hand, increasing the number of orbitals mostly favors spin fluctuations, although it might lead to a competition between charge and spin fluctuations. 
In turn, the spin and charge fluctuations are directly associated with the spin multiplicity of the superconducting state.
In general, spin fluctuations favor spin-singlet superconductivity and charge fluctuations favor spin-triplet superconductivity.

Away from the particle-hole ordered state, the fluctuations can trigger a finite superconducting critical temperature by two distinct mechanisms: ($i$) a $\boldsymbol{q}=0$ pairing vertex peak that requires \textit{attractive} interactions to generate a finite pairing strength \cite{berkEffectFerromagneticSpin1966}, and ($ii$) a finite $\boldsymbol{q}\neq0$ peak that increases the pairing strength even if the interactions are purely \textit{repulsive}.
The latter mechanism can increase the pairing strength enough to generate strong coupling superconductivity.

\ack
The authors thank Thereza A. Soares for manuscript proofreading. L.B.B. acknowledges financial support from Coordenação de Aperfeiçoamento de Pessoal de Nível Superior – Brasil (CAPES) – Finance Code 001. L.G.G.V.D.S. acknowledges financial support from CNPq (grants No. 309789/2020-6 and 312622/2023-6), and FAPESP (grant No. 2022/15453-0).




\begin{appendix}
\section{Interaction matrices in matrix-RPA theory}
\label{Sec:matrices}
In Sec.~\ref{sec:model}, we have shown the interaction matrices for general multiorbital and multisublattice systems with Hubbard-like and exchange-like interactions.
In this supplementary material, we present the procedure to derive those matrices, which we believe are of interest for new researches in the field, which are interested in using the matrix-RPA method.

Modeling Hubbard-like and exchange-like interactions requires one imposing to equation~\eqref{eq:Hgeneral} only inter-site interactions with density-density terms, and sublattice structure degrees of freedom.
The specific Hamiltonian then reads
\begin{equation}
\begin{split}
    H_{\text{int}}=&\sum_{ij}\sum_{\alpha\beta,s\sigma}\sum_{m_{\alpha\beta}}U_{m_{\alpha\beta}}n_{i\alpha s\uparrow}n_{j\beta s\downarrow} \\
    & +\sum_{ij}\sum_{\alpha\beta,p<s}\sum_{m_{\alpha\beta}}U'_{m_{\alpha\beta}}n_{i\alpha p}n_{j\beta s} \\
    &+\sum_{ij}\sum_{\alpha\beta,s<p,\sigma\sigma'}\sum_{m_{\alpha\beta}}J_{m_{\alpha\beta}}c_{i\alpha s\sigma}^{\dagger}c_{j\beta p\sigma'}^{\dagger}c_{j\alpha s\sigma'}c_{i \beta p\sigma}\\
	& +\sum_{ij}\sum_{\alpha\beta,s\neq p,\sigma\sigma'}\sum_{m_{\alpha\beta}}J'_{m_{\alpha\beta}}c_{i\alpha a\uparrow}^{\dagger}c_{j\beta p\downarrow}^{\dagger}c_{j\alpha s\downarrow}c_{i\beta p\uparrow}.
\end{split}
\label{eq:Ham_specific}
\end{equation}
Here, $s,p,t,d$ are orbital indices, $\sigma/\sigma'$ spin indices, $\alpha/\beta$ are sublattice indices, and $\{m_{\alpha\beta}\}$ denotes all the $w$ nearest-neighbors considered in the interaction ($m_{\alpha\beta}=0$ denotes onsite sublattice structure).
Also, $n_{i\alpha s} = n_{i\alpha s\uparrow} + n_{i\alpha s\downarrow}$, where $n_{i\alpha s\uparrow} = c^{\dagger}_{i\alpha s\uparrow}c_{i\alpha s\uparrow}$ is the sublattice and orbital density operator.

Next, we expand the orbital summation in equation~\eqref{eq:Hgeneral} noticing that a conservation of spin condition in present, namely $\sigma_{1}=\sigma_{4}$ and $\sigma_{2}=\sigma_{3}$, or $\sigma_{1}=\sigma_{3}$ and $\sigma_{2}=\sigma_{4}$ should be the only remaining elements. 
%
%
%
In addition, we to separate spin and charge terms [equation~\eqref{eq:charg_spin_parts}] using the following identity of Pauli matrices \cite{graserNeardegeneracySeveralPairing2009}:
\begin{equation}
    \boldsymbol{\sigma}_{\sigma_{1}\sigma_{4}}\cdot\boldsymbol{\sigma}_{\sigma_{2}\sigma_{3}}=2\delta_{\sigma_{1}\sigma_{3}}\delta_{\sigma_{2}\sigma_{4}}-\delta_{\sigma_{1}\sigma_{4}}\delta_{\sigma_{2}\sigma_{3}}
\end{equation}
to define the spin part of the interaction Hamiltonian.
In its turn, the charge component should resemble terms of the type
\begin{equation}
    \begin{split}
        \sum_{i\neq j}n_{i\alpha}n_{j\beta}	&= \sum_{i\neq j}\left(\sum_{s\sigma}c_{i\alpha s\sigma}^{\dagger}c_{i\alpha s\sigma}\right)\left(\sum_{p\sigma'}c_{j\beta p\sigma'}^{\dagger}c_{j\beta p\sigma'}\right)\\
	&=\sum_{i\neq j}\sum_{sp,\sigma\sigma'}c_{i\alpha s\sigma}^{\dagger}c_{i\alpha s\sigma}c_{j\beta p\sigma'}^{\dagger}c_{j\beta p,\sigma'}\\
	&=\sum_{i\neq j}\sum_{sp,\sigma\sigma'}c_{i\alpha s\sigma}^{\dagger}c_{j\beta p,\sigma'}^{\dagger}c_{j\beta p\sigma'}c_{i\alpha s\sigma},
    \end{split}
\end{equation}
that is, the charge component shows spin indices $\delta_{\sigma_{1}\sigma_{4}}\delta_{\sigma_{2}\sigma_{3}}$.
After some manipulation, equation~\eqref{eq:charg_spin_parts} becomes \cite{graserNeardegeneracySeveralPairing2009}
\begin{equation}
\begin{split}
    V_{\sigma_{1}\sigma_{2}\sigma_{3}\sigma_{4}}^{\alpha s\tilde{\alpha} p,\beta t\tilde{\beta} d} = 
    &-\mathcal{V}^{\alpha s\tilde{\alpha} p}_{\beta t\tilde{\beta} d}\delta_{\sigma_{1}\sigma_{3}}\delta_{\sigma_{2}\sigma_{4}}\\
    &+ \frac{1}{2} \left(\mathcal{U}^{\alpha s\tilde{\alpha} p}_{\beta t\tilde{\beta} d} + \mathcal{V}^{\alpha s\tilde{\alpha} p}_{\beta t\tilde{\beta} d} \right)\delta_{\sigma_{2}\sigma_{3}}\delta_{\sigma_{1}\sigma_{4}}.
\end{split}
\label{eq:charge_spin_sm}
\end{equation}
Now, having the interactions separated into the two main electron properties for condensed matter, charge and spin parts, we move to fing the the charge $\hat{\mathcal{U}}$ and spin $\hat{\mathcal{V}}$ interaction matrices.
Expanding the summation over orbitals in the interaction Hamiltonian of equation~\eqref{eq:Hgeneral}, one obtains
\begin{align}
    H_{\text{int}}&=\sum_{ij}\sum_{\alpha\beta,\sigma_{l}}{\text{\c{c}}^{\alpha\beta}_{ij}}\notag\\
    &\;\;\Big(\sum_{s}V_{\sigma_{1}\sigma_{2}\sigma_{3}\sigma_{4},ij}^{\alpha s\alpha s,\beta s\beta s}c_{i\alpha s\sigma_{1}}^{\dagger}c_{j\beta s\sigma_{2}}^{\dagger}c_{j\beta s\sigma_{3}}c_{i\alpha s\sigma_{4}}\label{eq:Hrow1} \\
    &+\sum_{s\neq p}V_{\sigma_{1}\sigma_{2}\sigma_{3}\sigma_{4},ij}^{\alpha s\alpha p,\beta s\beta p}c_{i\alpha s\sigma_{1}}^{\dagger}c_{j\beta p\sigma_{2}}^{\dagger}c_{j\beta s\sigma_{3}}c_{i\alpha p\sigma_{4}}\label{eq:Hrow2}\\
    &+\sum_{s\neq p}V_{\sigma_{1}\sigma_{2}\sigma_{3}\sigma_{4},ij}^{\alpha s\alpha p,\beta p\beta s}c_{i\alpha s\sigma_{1}}^{\dagger}c_{j\beta p\sigma_{2}}^{\dagger}c_{j\beta p\sigma_{3}}c_{i\alpha s\sigma_{4}}\label{eq:Hrow3}\\
    &+\sum_{s\neq p}V_{\sigma_{1}\sigma_{2}\sigma_{3}\sigma_{4},ij}^{\alpha s\alpha s,\beta p\beta p}c_{i\alpha s\sigma_{1}}^{\dagger}c_{j\beta s\sigma_{2}}^{\dagger}c_{j\beta p\sigma_{3}}c_{i\alpha p\sigma_{4}}\Big)\label{eq:Hrow4}
\end{align}
Any other orbital terms would give null matrix elements also because of Pauli exclusion principle.

The procedure to obtain the spin and charge interaction matrices reduces to identify the terms in Eqs.~\eqref{eq:Hrow1}-\eqref{eq:Hrow4} with the respective term in equation~\eqref{eq:Ham_specific}.
Here, we will show as an example how to obtain the intraorbital elements [equation~\eqref{eq:Hrow1}].

equation~\eqref{eq:Hrow1} shows same-orbital four-point interactions, similarly to the usual Hubbard Hamiltonian when $\alpha=\beta$, $H_{\text{Hubbard}}=\frac{U}{2}\sum_{i,\alpha s,\sigma_{1}\neq\sigma_{2}}n_{i\alpha s\sigma_{1}}n_{i\alpha s\sigma_{2}}=\frac{U}{2}\sum_{i\alpha s\sigma_{1}\neq\sigma_{2}}c_{i\alpha c\sigma_{1}}^{\dagger}c_{i\alpha c\sigma_{1}}c_{i\alpha c\sigma_{2}}^{\dagger}c_{i\alpha c\sigma_{2}}$.
Taking the spin into consideration and using the definition of equation~\eqref{eq:charge_spin_sm}, one obtains
%
\begin{equation}
    \begin{split}
        V_{\sigma_{1}\sigma_{2}\sigma_{3}\sigma_{4},ij}^{\alpha s\alpha s,\beta s\beta s}c_{i\alpha s,\sigma_{1}}^{\dagger}c_{j\beta s\sigma_{2}}^{\dagger}c_{j\beta s\sigma_{3}}c_{i\alpha s\sigma_{4}} =   \\
        \left[-\mathcal{V}^{\alpha s\alpha s}_{\alpha s\alpha s} \delta_{\sigma_{1}\sigma_{3}}\delta_{\sigma_{2}\sigma_{4}}+\frac{1}{2}\left(\mathcal{U}^{\alpha s\alpha s}_{\alpha s\alpha s} +\mathcal{V}^{\alpha s\alpha s}_{\alpha s\alpha s}\right) \delta_{\sigma_{2}\sigma_{3}}\delta_{\sigma_{1}\sigma_{4}}\right]  \\
        \times c_{i\alpha s\sigma_{1}}^{\dagger}c_{j\beta s\sigma_{2}}^{\dagger}c_{j\beta s\sigma_{3}}c_{i\alpha s\sigma_{4}}\\
         = -\mathcal{V}^{\alpha s\alpha s}_{\alpha s\alpha s} c_{i\alpha s\sigma_{1}}^{\dagger}c_{j\beta s\sigma_{2}}^{\dagger}c_{j\beta s\sigma_{1}}c_{i\alpha s\sigma_{2}} \\
         +\frac{1}{2}\left(\mathcal{U}^{\alpha s\alpha s}_{\alpha s\alpha s} +\mathcal{V}^{\alpha s\alpha s}_{\alpha s\alpha s}\right) c_{i\alpha s\sigma_{1}}^{\dagger}c_{j\beta s\sigma_{2}}^{\dagger}c_{j\beta s\sigma_{2}}c_{i\alpha s\sigma_{1}}\\
         = -\mathcal{V}^{\alpha s\alpha s}_{\alpha s\alpha s} c_{i\alpha s\sigma_{1}}^{\dagger}c_{j\beta s\sigma_{2}}^{\dagger}c_{j\beta s\sigma_{1}}c_{i\alpha s\sigma_{2}} \\
         +\frac{1}{2}\left(\mathcal{U}^{\alpha s\alpha s}_{\alpha s\alpha s} +\mathcal{V}^{\alpha s\alpha s}_{\alpha s\alpha s}\right) c_{i\alpha s\sigma_{1}}^{\dagger}c_{i\alpha s\sigma_{1}}c_{j\beta s\sigma_{2}}^{\dagger}c_{j\beta s\sigma_{2}} \\
         = -\mathcal{V}^{\alpha s\alpha s}_{\alpha s\alpha s} c_{i\alpha s\sigma_{1}}^{\dagger}c_{j\beta s\sigma_{2}}^{\dagger}c_{j\beta s\sigma_{1}}c_{i\alpha s\sigma_{2}} \\
         +\frac{1}{2}\left(\mathcal{U}^{\alpha s\alpha s}_{\alpha s\alpha s} +\mathcal{V}^{\alpha s\alpha s}_{\alpha s\alpha s}\right) n_{i\alpha s\sigma_{1}}n_{j\beta s\sigma_{2}}.
    \end{split}
\end{equation}
%

The next step is to write the density-density term in momentum space.
First, let us consider the transformation $c_{j\alpha s\sigma}^{\dagger}=\sum_{\boldsymbol{k}}e^{-i\boldsymbol{k}\cdot\boldsymbol{r}_{j\alpha}}c_{\boldsymbol{k}\alpha s,\sigma}^{\dagger}/\sqrt{N}$ (with $N$ being the  the number of lattice sites) in the first term of the above equality:
\begin{equation}
    \begin{split}
         \sum_{\substack{i \\ m_{\alpha\beta},l} } \mathcal{V}^{\alpha s\alpha s}_{\alpha s\alpha s} c_{i\alpha s\sigma_{1}}^{\dagger}c_{j=(i+m_{\alpha\beta}\beta,l) s\sigma_{2}}^{\dagger}c_{j=(i+m_{\alpha\beta}\beta,l) s\sigma_{1}}c_{i\alpha s\sigma_{2}}  \\
        =  \frac{1}{N^{2}}  \sum_{\substack{i \\ m_{\alpha\beta},l \\ \boldsymbol{k}_{1},\boldsymbol{k}_{2} \\ \boldsymbol{k}_{3},\boldsymbol{k}_{4} }   } 
        \mathcal{V}^{\alpha s\alpha s}_{\alpha s\alpha s} e^{-i\boldsymbol{r}_{i\alpha}\cdot(\boldsymbol{k}_{1}-\boldsymbol{k}_{4})}e^{-i\boldsymbol{r}_{(i+m_{\alpha\beta},l)\beta}\cdot(\boldsymbol{k}_{2}-\boldsymbol{k}_{3})} \\
         \times c_{\boldsymbol{k}_{1}\alpha s\sigma_{1}}^{\dagger}c_{\boldsymbol{k}_{2}\beta s\sigma_{2}}^{\dagger}c_{\boldsymbol{k}_{3}\beta s\sigma_{1}}c_{\boldsymbol{k}_{4}\alpha s\sigma_{2}} \\
        =  \frac{1}{N^{2}} \sum_{\substack{m_{\alpha\beta}, l \\ \boldsymbol{k}_{1},\boldsymbol{k}_{2} \\ \boldsymbol{k}_{3},\boldsymbol{k}_{4} }   }
        e^{-i\boldsymbol{r}_{i\alpha}\cdot(\boldsymbol{k}_{1}+\boldsymbol{k}_{2}-(\boldsymbol{k}_{3}+\boldsymbol{k}_{4}))}
         e^{-i\boldsymbol{\delta}_{m_{\alpha\beta},l}\cdot(\boldsymbol{k}_{2}-\boldsymbol{k}_{3})} \\
         \times  c_{\boldsymbol{k}_{1}\alpha s\sigma_{1}}^{\dagger}c_{\boldsymbol{k}_{2}\beta s\sigma_{2}}^{\dagger}c_{\boldsymbol{k}_{3}\beta s\sigma_{1}}c_{\boldsymbol{k}_{4}\alpha s\sigma_{2}} \\
         =\frac{1}{N} \sum_{\substack{m_{\alpha\beta}, l \\ \boldsymbol{q},\boldsymbol{k}_{1} \\ \boldsymbol{k}_{3},\boldsymbol{k}_{4} } }
         \mathcal{V}^{\alpha s\alpha s}_{\alpha s\alpha s} e^{-i\boldsymbol{\delta}_{m_{\alpha\beta},l}\cdot\boldsymbol{q}}\\
         \times c_{\boldsymbol{k}_{1}\alpha s\sigma_{1}}^{\dagger}c_{\boldsymbol{k}_{3}+\boldsymbol{q}\beta s\sigma_{2}}^{\dagger}c_{\boldsymbol{k}_{3}\beta s\sigma_{1}}c_{\boldsymbol{k}_{4}\alpha s\sigma_{2}}\delta_{\boldsymbol{k}_{1}+\boldsymbol{k}_{3}+\boldsymbol{q},\boldsymbol{k}_{3}+\boldsymbol{k}_{4}}\\
         = \frac{1}{N} \sum_{\substack{m_{\alpha\beta}, l \\ \boldsymbol{q} \\ \boldsymbol{k}_{3},\boldsymbol{k}_{4} } }
\mathcal{V}^{\alpha s\alpha s}_{\alpha s\alpha s} e^{-i\boldsymbol{\delta}_{m_{\alpha\beta},l}\cdot\boldsymbol{q}}\\
         \times c_{\boldsymbol{k}_{4}-\boldsymbol{q}\alpha s\sigma_{1}}^{\dagger}c_{\boldsymbol{k}_{4}\alpha s\sigma_{2}}c_{\boldsymbol{k}_{3}+\boldsymbol{q}\beta s\sigma_{2}}^{\dagger}c_{\boldsymbol{k}_{3}\beta s\sigma_{1}},
    \end{split}
\end{equation}
where we used that $\boldsymbol{r}_{j\alpha}=\boldsymbol{r}_{(i+m_{\alpha\beta},l)\beta}=\sum_l\boldsymbol{r}_{i\alpha}+\boldsymbol{\delta}_{m_{\alpha\beta},l}$ ($l$ indexing each $m_{\alpha\beta}$-th nearest-neighbor) and the identity $\sum_{j}e^{-i(\boldsymbol{k}-\boldsymbol{k}')\cdot\boldsymbol{r}_{j\alpha}}=N\delta_{\boldsymbol{k},\boldsymbol{k}'}$.
Also, it was convenient to define $\boldsymbol{k}_{2}-\boldsymbol{k}_{3}=\boldsymbol{q}\Rightarrow\boldsymbol{k}_{2}=\boldsymbol{k}_{3}+\boldsymbol{q}$, while the Dirac delta $\delta_{\boldsymbol{k}_{1}+\boldsymbol{k}_{3}+\boldsymbol{q},\boldsymbol{k}_{3}+\boldsymbol{k}_{4}}$ imposed $\boldsymbol{k}_{1}+\boldsymbol{q}=\boldsymbol{k}_{4}\Rightarrow\boldsymbol{k}_{1}=\boldsymbol{k}_{4}-\boldsymbol{q}$.
The above term even in momentum space cannot be realized if $\alpha\neq\beta$ since there is exchange of spin in the creation and annihilation operators.
Therefore, we must set $\mathcal{V}^{\alpha s\alpha s}_{\alpha s\alpha s}=0$ for $\alpha\neq\beta$.
However, the spin part is non-zero when $\alpha=\beta$ because the above terms can be reorganized in density terms, as seen below.

Following the same procedure in the second term of the above equality, which is already a density-density term, results in
\begin{equation}
    \begin{split}
        \sum_i \sum_{m_{\alpha\beta},l} \frac{1}{2}\left(\mathcal{U}^{\alpha s\alpha s}_{\alpha s\alpha s} +\mathcal{V}^{\alpha s\alpha s}_{\alpha s\alpha s}\right)&\\
        \times c_{i\alpha s\sigma_{1}}^{\dagger}c_{i\alpha s\sigma_{1}}c_{j=(i+m_{\alpha\beta},l)\beta s\sigma_{2}}^{\dagger}c_{j=(i+m_{\alpha\beta},l)\beta s\sigma_{2}}&\\
        = \frac{1}{N}\sum_{m_{\alpha\beta},l}\sum_{\boldsymbol{q}}\frac{1}{2}\left(\mathcal{U}^{\alpha s\alpha s}_{\alpha s\alpha s} +\mathcal{V}^{\alpha s\alpha s}_{\alpha s\alpha s}\right)e^{-i\boldsymbol{\delta}_{m_{\alpha\beta},l}\cdot\boldsymbol{q}}&\\
        n_{\alpha s\sigma_{1}}(-\boldsymbol{q})n_{\beta s\sigma_{2}}(\boldsymbol{q}).&
    \end{split}
\end{equation}
At this point we can identify this last term with intersite interaction $\frac{1}{2}\sum_{\sigma_{g}}\sum_{ij}\sum_{\alpha\beta}\sum_{m_{\alpha\beta}}U_{m_{\alpha\beta}}n_{i\alpha s\sigma_{1}}n_{j\beta s\sigma_{2}}$.
For $\alpha\neq\beta$, $\mathcal{V}^{\alpha s\alpha s}_{\alpha s\alpha s}=0$, so $\mathcal{U}^{\alpha s\alpha s}_{\alpha s\alpha s} = 2\cdot2\cdot\sum_{m_{\alpha\beta}}U_{m_{\alpha\beta}}e^{i\boldsymbol{\delta}_{m_{\alpha\beta},l}\cdot\boldsymbol{q}}/2 = 2\cdot\sum_{m_{\alpha\beta},l}U_{m_{\alpha\beta}}e^{i\boldsymbol{\delta}_{m_{\alpha\beta},l}\cdot\boldsymbol{q}} = U_{\alpha\beta}$ (one of the factors of 2 comes from the matrix element we just dissected, while the other one comes from ${\text{\c{c}}^{\alpha\beta}_{ij}}=1/2$ for $\alpha\neq\beta$).
When $\alpha=\beta$, the term $\mathcal{V}^{\alpha s\alpha s}_{\alpha s\alpha s}$ will be non-zero, as well as $\mathcal{U}^{\alpha s\alpha s}_{\alpha s\alpha s}$, and in momentum space they will appear with a form factor $\mathcal{U}^{\alpha s\alpha s}_{\alpha s\alpha s} = \mathcal{V}^{\alpha s\alpha s}_{\alpha s\alpha s} = \sum_{m_{\alpha\beta},l}U_{m_{\alpha\alpha}}e^{i\boldsymbol{\delta}_{m_{\alpha\alpha},l}\cdot\boldsymbol{q}} = U_{\alpha\alpha}$, which is obtained by comparing the above results with equation~\eqref{eq:Hrow1}.

The other matrix elements can be obtained by a similar procedure.

\section{Derivation of bounds for eigenvalues}
\label{Sec:trace_bounds}
In order to keep track of the spin-charge competition we evaluate the main eigenvalue of each of these equations and compare them. 
It is not enough to subtract the equations once we would not keep track of the main eigenvalues in this way. 
We now present the result of Theorem 2.1 of the work \cite{wolkowiczBoundsEigenvaluesUsing1980} stating that if an $n_{\mathcal{A}}\times n_{\mathcal{A}}$ matrix $\mathcal{A}$ complex matrix $\mathcal{A}$ has reals eigenvalues and minimum eigenvalue $\lambda_{\text{min}}^{\mathcal{A}}$ and maximum eigenvalue $\lambda_{\text{max}}^{\mathcal{A}}$, and letting 
\begin{align}
    m_{\mathcal{A}} & = \text{tr}\mathcal{A}/n_{\mathcal{A}}, \\
    s_{\mathcal{A}}^{2}	&=\text{tr}\mathcal{A}^{2}/n_{\mathcal{A}}-m_{\mathcal{A}}^{2},
\end{align}
then
\begin{align}
    m_{\mathcal{A}}-s_{\mathcal{A}}(n_{\mathcal{A}}-1)^{1/2}\leq&\lambda_{\text{min}}^{\mathcal{A}}\leq m_{\mathcal{A}}-s_{\mathcal{A}}/(n_{\mathcal{A}}-1)^{1/2}, \\
    m_{\mathcal{A}}+s_{\mathcal{A}}/(n_{\mathcal{A}}-1)^{1/2}	\leq&\lambda_{\text{max}}^{\mathcal{A}}\leq m_{\mathcal{A}}+s_{\mathcal{A}}(n_{\mathcal{A}}-1)^{1/2}.
\end{align}
These are bounds for the maximum and minimum eigenvalues of Hermitian matrices. 
They are particularly useful for analyzing the Stoner criteria, because in the diagonal approximation \cite{kemperSensitivitySuperconductingState2010} the product of $\hat{\chi}_{0}$ and each of the interaction matrices is Hermitian if $\chi_{\alpha\alpha}^{sp}=\chi_{\alpha\alpha}^{ps}$ for any pair $p,s$ (this is our main approximation for the next analysis), because the interaction matrices ($\mathcal{\hat{U}},\hat{\mathcal{V}}$) are Hermitian and, in this way, $\hat{\chi}_{0}$ is Hermitian as well. 
Next, we derive the boundaries of the maximum eigenvalue of the charge and spin Stoner criteria and, at the end, compare both of them.

We start with the spin Stoner criterion. 
Under the diagonal approximation, we obtain blocks
\begin{align}
    \hat{\mathcal{V}}\hat{\chi}	&=
    \begin{bmatrix}
    \hat{\mathcal{V}}_{ss}\hat{\chi}_{ss} & \hat{0} & \hat{0} & \hat{0}\\
    \hat{0} & \hat{\mathcal{V}}_{sp}\hat{\chi}_{sp} & \hat{0} & \hat{0}\\
    \hat{0} & \hat{0} & \hat{0} & \hat{0}\\
    \hat{0} & \hat{0} & \hat{0} & \hat{0}
    \end{bmatrix}, \\
    \hat{\mathcal{V}}_{ss}\hat{\chi}_{ss} &= 
    \begin{bmatrix}
    \hat{\mathcal{V}}_{AA}^{ss}\hat{\chi}_{AA}^{ss} & \hat{0} & \hat{0} & \cdots\\
    \hat{0} & \hat{\mathcal{V}}_{BB}^{ss}\hat{\chi}_{BB}^{ss} & \hat{0} & \cdots\\
    \hat{0} & \hat{0} & \hat{\mathcal{V}}_{CC}^{ss}\hat{\chi}_{BB}^{ss} & \cdots\\
    \vdots & \vdots & \vdots & \ddots
    \end{bmatrix}, \\
    \hat{\mathcal{V}}_{sp}\hat{\chi}_{sp}	& =\text{diag}(\hat{\mathcal{V}}^{sp}\hat{\chi}^{sp},...,\hat{\mathcal{V}}^{sp}\hat{\chi}^{sp}),\\
    \hat{\mathcal{V}}^{sp}\hat{\chi}^{sp} &=
    \begin{bmatrix}
    \hat{\mathcal{V}}_{AA}^{sp}\hat{\chi}_{AA}^{sp} & \hat{0} & \hat{0} & \cdots\\
    \hat{0} & \hat{\mathcal{V}}_{BB}^{sp}\hat{\chi}_{BB}^{sp} & \hat{0} & \cdots\\
    \hat{0} & \hat{0} & \hat{\mathcal{V}}_{CC}^{sp}\hat{\chi}_{CC}^{sp} & \cdots\\
    \vdots & \vdots & \vdots & \ddots
    \end{bmatrix}, \\
    \hat{\mathcal{V}}_{\alpha\alpha}^{ss}\hat{\chi}_{\alpha\alpha}^{ss}	&=\chi_{\alpha\alpha}^{ss}
    \begin{bmatrix}
    U_{\alpha\alpha} & J_{\alpha\alpha} & J_{\alpha\alpha} & \cdots\\
    J_{\alpha\alpha} & U_{\alpha\alpha} & J_{\alpha\alpha} & \cdots\\
    J_{\alpha\alpha} & J_{\alpha\alpha} & U_{\alpha\alpha} & \cdots\\
    \vdots & \vdots & \vdots & \ddots
    \end{bmatrix},\\
    \hat{\mathcal{V}}_{\alpha\alpha}^{sp}\hat{\chi}_{\alpha\alpha}^{sp}
    &=\chi_{\alpha\alpha}^{sp}
    \begin{bmatrix}
    J'_{\alpha\alpha} & U'_{\alpha\alpha}\\
    U'_{\alpha\alpha} & J'_{\alpha\alpha}
    \end{bmatrix}.
\end{align}
Using $n_{\hat{\mathcal{V}}_{\alpha\alpha}^{sp}\hat{\chi}_{\alpha\alpha}^{sp}}=2$, $\text{tr}\hat{\mathcal{V}}_{\alpha\alpha}^{sp}\hat{\chi}_{\alpha\alpha}^{sp}=\chi_{\alpha\alpha}^{sp}2J'_{\alpha\alpha}$, and $\text{tr}(\hat{\mathcal{V}}_{\alpha\alpha}^{sp}\hat{\chi}_{\alpha\alpha}^{sp})^{2}=(\chi_{\alpha\alpha}^{sp})^{2}2\left((J'_{\alpha\alpha})^2+(U'_{\alpha\alpha})^{2} \right)$, we obtain parameters
\begin{equation}
    \begin{split}
        m_{\hat{\mathcal{V}}_{\alpha\alpha}^{sp}\hat{\chi}_{\alpha\alpha}^{sp}}	&=(\chi_{\alpha\alpha}^{sp})2J'_{\alpha\alpha}/2=\chi_{\alpha\alpha}^{sp}J'_{\alpha\alpha}, \\
    s_{\hat{\mathcal{V}}_{\alpha\alpha}^{sp}\hat{\chi}_{\alpha\alpha}^{sp}}^{2}	&=(\chi_{\alpha\alpha}^{sp})^{2}2\left((J'_{\alpha\alpha})^{2}+(U'_{\alpha\alpha})^{2} \right)/2-m_{\hat{\mathcal{V}}_{\alpha\alpha}^{sp}\hat{\chi}_{\alpha\alpha}^{sp}}^{2} \\
    &=(\chi_{\alpha\alpha}^{sp})^{2}\left((J'_{\alpha\alpha})^{2}+(U'_{\alpha\alpha})^{2}-(J'_{\alpha\alpha})^{2}\right) \\
    \Rightarrow s_{\hat{\mathcal{V}}_{\alpha\alpha}^{sp}\hat{\chi}_{\alpha\alpha}^{sp}}	&=\chi_{\alpha\alpha}^{sp}U'_{\alpha\alpha}.
    \end{split}
\end{equation}
The boundaries for the eigenvalues are
\begin{equation}
    \begin{split}
        \chi_{\alpha\alpha}^{sp}(J'_{\alpha\alpha}+U'_{\alpha\alpha})	\leq\lambda_{\text{max}}^{\hat{\mathcal{V}}_{\alpha\alpha}^{sp}\hat{\chi}_{\alpha\alpha}^{sp}}\leq\chi_{\alpha\alpha}^{sp}(J'_{\alpha\alpha}+U'_{\alpha\alpha}) \\
    \Rightarrow	\lambda_{\text{max}}^{\hat{\mathcal{V}}_{\alpha\alpha}^{sp}\hat{\chi}_{\alpha\alpha}^{sp}}=\chi_{\alpha\alpha}^{sp}(J'_{\alpha\alpha}+U'_{\alpha\alpha}).
    \end{split}
\end{equation}
It is trivial to compute the eigenvalue of a $2\times2$ matrix but here we show an example where the theorem works well to determine the main eigenvalue. 
Now, we move to $\hat{\mathcal{V}}_{\alpha\alpha}^{ss}\hat{\chi}_{\alpha\alpha}^{ss}$, where $n_{\hat{\mathcal{V}}_{\alpha\alpha}^{ss}\hat{\chi}_{\alpha\alpha}^{ss}}=N_{o}, \text{tr}\hat{\mathcal{V}}_{\alpha\alpha}^{ss}\hat{\chi}_{\alpha\alpha}^{ss}=\chi_{\alpha\alpha}^{ss}N_{o}U_{\alpha\alpha}$, and $\text{tr}(\hat{\mathcal{V}}_{\alpha\alpha}^{ss}\hat{\chi}_{\alpha\alpha}^{ss})^{2}=(\chi_{\alpha\alpha}^{ss})^{2}(U_{\alpha\alpha}^{2}+J_{\alpha\alpha}^{2}(N_{o}-1))N_{o}$. 
We, therefore, obtain parameters
\begin{equation}
    \begin{split}
        m_{\hat{\mathcal{V}}_{\alpha\alpha}^{ss}\hat{\chi}_{\alpha\alpha}^{ss}}	=\chi_{\alpha\alpha}^{ss}N_{o}U_{\alpha\alpha}/N_{o}=\chi_{\alpha\alpha}^{ss}U_{\alpha\alpha}, &\\
        s_{\hat{\mathcal{V}}_{\alpha\alpha}^{ss}\hat{\chi}_{\alpha\alpha}^{ss}}^{2}	=(\chi_{\alpha\alpha}^{ss})^{2}(U_{\alpha\alpha}^{2}+J_{\alpha\alpha}^{2}(N_{o}-1))N_{o}/N_{o}&\\
        -m_{\hat{\mathcal{V}}_{\alpha\alpha}^{ss}\hat{\chi}_{\alpha\alpha}^{ss}}^{2} &\\
	  =(\chi_{\alpha\alpha}^{ss})^{2}(U_{\alpha\alpha}^{2}+J_{\alpha\alpha}^{2}(N_{o}-1)-U_{\alpha\alpha}^{2}) &\\
        \Rightarrow s_{\hat{\mathcal{V}}_{\alpha\alpha}^{sp}\hat{\chi}_{\alpha\alpha}^{sp}}	=\chi_{\alpha\alpha}^{ss}J_{\alpha\alpha}(N_{o}-1)^{1/2}.&
    \end{split}
\end{equation}
Thus, the boundaries are 
\begin{equation}
    \chi_{\alpha\alpha}^{ss}(U_{\alpha\alpha}+J_{\alpha\alpha})	\leq\lambda_{\text{max}}^{\hat{\mathcal{V}}_{\alpha\alpha}^{ss}\hat{\chi}_{\alpha\alpha}^{ss}}\leq\chi_{\alpha\alpha}^{ss}(U_{\alpha\alpha}+J_{\alpha\alpha}(N_{o}-1)).
\end{equation}
Notice the upper boundary here grows with the number of orbitals. 
By a simple inspection of the typical values of the non-interacting susceptibility, we conclude that it is more likely that $\text{max}\alpha_{s}=\max_{\alpha}\lambda_{\text{max}}^{\hat{\mathcal{V}}_{\alpha\alpha}^{ss}\hat{\chi}_{\alpha\alpha}^{ss}}>\max_{\alpha}\lambda_{\text{max}}^{\hat{\mathcal{V}}_{\alpha\alpha}^{sp}\hat{\chi}_{\alpha\alpha}^{sp}}$, meaning the homogeneous elements dominate the spin fluctuations.

Next, we move to the charge channel. 
On the charge channel, however, the previous diagonal bare susceptibility approximation is not enough for keeping the product of matrices Hermitian. 
The other assumption must be that the susceptibility does not show preference for any specific sublattice site, that is $\hat{\chi}_{\alpha\beta}^{ss}=\hat{\chi}_{\alpha\beta}^{pp}$ and $\hat{\chi}_{\alpha\beta}^{sp}=\hat{\chi}_{\alpha\beta}^{ps}$ for any $\alpha,\beta,s,p$. 
Although physically this may seem a reasonable assumption, it might not be realistic in systems with fractional dimension, such as thick graphene samples where the outer, surface sublattice sites have higher density of states than the inner sites. 
Under this approximation, the charge Stoner parameter depends on the main eigenvalue of the following matrix
\begin{equation}
    \hat{\mathcal{U}}\hat{\chi}=\begin{bmatrix}\hat{\mathcal{U}}_{ss}\hat{\chi}_{ss} & \hat{0} & \hat{0} & \hat{0}\\
    \hat{0} & \hat{\mathcal{U}}_{sp}\hat{\chi}_{sp} & \hat{0} & \hat{0}\\
    \hat{0} & \hat{0} & \hat{0} & \hat{0}\\
\hat{0} & \hat{0} & \hat{0} & \hat{0}
    \end{bmatrix},
\end{equation}
\begin{equation}
\hat{\mathcal{U}}_{ss}\hat{\chi}_{ss}=\begin{bmatrix}\hat{\mathcal{U}}_{AA}^{ss}\hat{\chi}_{\alpha\alpha}^{ss} & \hat{\mathcal{U}}_{AB}^{ss}\hat{\chi}_{\alpha\alpha}^{ss} & \hat{\mathcal{U}}_{AC}^{ss}\hat{\chi}_{\alpha\alpha}^{ss} & \cdots\\
    \hat{\mathcal{U}}_{AB}^{ss\dagger}\hat{\chi}_{\alpha\alpha}^{ss} & \hat{\mathcal{U}}_{BB}^{ss}\hat{\chi}_{\alpha\alpha}^{ss} & \hat{\mathcal{U}}_{BC}^{ss}\hat{\chi}_{\alpha\alpha}^{ss} & \cdots\\
    \hat{\mathcal{U}}_{AC}^{ss\dagger}\hat{\chi}_{\alpha\alpha}^{ss} & \hat{\mathcal{U}}_{BC}^{ss\dagger}\hat{\chi}_{\alpha\alpha}^{ss} & \hat{\mathcal{U}}_{CC}^{ss}\hat{\chi}_{\alpha\alpha}^{ss} & \cdots\\
    \vdots & \vdots & \vdots & \ddots \\
    \end{bmatrix},
\end{equation}
\begin{equation}
    \hat{\mathcal{U}}_{sp}\hat{\chi}_{sp}=\text{diag}(\hat{\mathcal{U}}^{sp}\hat{\chi}^{sp},...,\hat{\mathcal{U}}^{sp}\hat{\chi}^{sp}),
\end{equation}
\begin{equation}
\hat{\mathcal{U}}^{sp}\hat{\chi}^{sp}=\begin{bmatrix}\hat{\mathcal{U}}_{AA}^{sp}\hat{\chi}_{\alpha\alpha}^{sp} & \hat{\mathcal{U}}_{AB}^{sp}\hat{\chi}_{\alpha\alpha}^{sp} & \hat{\mathcal{U}}_{AC}^{sp}\hat{\chi}_{\alpha\alpha}^{sp} & \cdots\\
\hat{\mathcal{U}}_{AB}^{sp\dagger}\hat{\chi}_{\alpha\alpha}^{sp} & \hat{\mathcal{U}}_{BB}^{sp}\hat{\chi}_{\alpha\alpha}^{sp} & \hat{\mathcal{U}}_{BC}^{sp}\hat{\chi}_{\alpha\alpha}^{sp} & \cdots\\
\hat{\mathcal{U}}_{AC}^{sp\dagger}\hat{\chi}_{\alpha\alpha}^{sp} & \hat{\mathcal{U}}_{BC}^{sp\dagger}\hat{\chi}_{\alpha\alpha}^{sp} & \hat{\mathcal{U}}_{CC}^{sp}\hat{\chi}_{\alpha\alpha}^{sp} & \cdots\\
\vdots & \vdots & \vdots & \ddots
\end{bmatrix},
\end{equation}
\begin{equation}
\hat{\mathcal{U}}_{\alpha\alpha}^{ss}\hat{\chi}_{\alpha\alpha}^{ss}=\chi_{\alpha\alpha}^{ss}\begin{bmatrix}U_{\alpha\alpha} & N_{\alpha\alpha} & N_{\alpha\alpha} & \cdots\\
N_{\alpha\alpha} & U_{\alpha\alpha} & N_{\alpha\alpha} & \cdots\\
N_{\alpha\alpha} & N_{\alpha\alpha} & U_{\alpha\alpha} & \cdots\\
\vdots & \vdots & \vdots & \ddots
\end{bmatrix},
\end{equation}
\begin{equation}
\hat{\mathcal{U}}_{\alpha\alpha}^{sp}\hat{\chi}_{\alpha\alpha}^{sp}=\chi_{\alpha\alpha}^{sp}\begin{bmatrix}J'_{\alpha\alpha} & M_{\alpha\alpha}\\
M_{\alpha\alpha} & J'_{\alpha\alpha}
\end{bmatrix},
\end{equation}
\begin{equation}
\hat{\mathcal{U}}_{\alpha\beta}^{ss}\hat{\chi}_{\alpha\alpha}^{ss}=\chi_{\alpha\alpha}^{ss}\begin{bmatrix}U_{\alpha\beta} & U'_{\alpha\beta} & U'_{\alpha\beta} & \cdots\\
U'_{\alpha\beta} & U_{\alpha\beta} & U'_{\alpha\beta} & \cdots\\
U'_{\alpha\beta} & U'_{\alpha\beta} & U_{\alpha\beta} & \cdots\\
\vdots & \vdots & \vdots & \cdots
\end{bmatrix},
\end{equation}
\begin{equation}
\hat{\mathcal{U}}_{\alpha\beta}^{sp}\hat{\chi}_{\alpha\alpha}^{sp}=\chi_{\alpha\alpha}^{sp}\begin{bmatrix}J'_{\alpha\beta} & J_{\alpha\beta}\\
J_{\alpha\beta} & J'_{\alpha\beta}
\end{bmatrix}.
\end{equation}
Next, we derive the lower eigenvalues of each of these block matrices, because the Stoner criterion includes a minus sign. 
For simplicity, consider $J'_{\alpha\alpha}=J'_{\beta\beta},J_{\alpha\alpha}=J_{\beta\beta},M_{\alpha\alpha}=M_{\beta\beta},N_{\alpha\alpha}=N_{\beta\beta}$ for any $\alpha,\beta$. 
We start with the $2N_{s}\times2N_{s}$ $\hat{\mathcal{U}}^{sp}\hat{\chi}^{sp}$ blocks, where $n_{\hat{\mathcal{U}}^{sp}\hat{\chi}^{sp}}=2N_{s}$, $\text{tr}\hat{\mathcal{U}}^{sp}\hat{\chi}^{sp}=\chi_{\alpha\alpha}^{sp}N_{s}2J'_{\alpha\alpha}$, $\text{tr}(\hat{\mathcal{U}}^{sp}\hat{\chi}^{sp})^{2}=(\chi_{\alpha\alpha}^{sp})^{2}\left[(J'_{\alpha\alpha})^{2}+M_{\alpha\alpha}^{2}+\sum_{\alpha>\beta}(|J'_{\alpha\beta}|^{2}+|J_{\alpha\beta}|^{2})\right]2N_{s}$, resulting in parameters
\begin{equation}
    \begin{split}
        m_{\hat{\mathcal{U}}^{sp}\hat{\chi}^{sp}}	=\chi_{\alpha\alpha}^{sp}2N_{s}J'_{\alpha\alpha}/(2N_{s})=\chi_{\alpha\alpha}^{sp}J'_{\alpha\alpha},& \\
        s_{\hat{\mathcal{U}}^{sp}\hat{\chi}^{sp}}^{2}	=\left[(J'_{\alpha\alpha})^{2}+M_{\alpha\alpha}^{2}+\sum_{\alpha>\beta}(|J'_{\alpha\beta}|^{2}+|J_{\alpha\beta}|^{2})\right]&\\
        \times 2N_{s}/(2N_{s})(\chi_{\alpha\alpha}^{sp})^{2}-m_{\hat{\mathcal{U}}^{sp}\hat{\chi}^{sp}}^{2} &\\
	   =(\chi_{\alpha\alpha}^{sp})^{2}[M_{\alpha\alpha}^{2}+\sum_{\alpha>\beta}(|J'_{\alpha\beta}|^{2}+|J_{\alpha\beta}|^{2})].&
    \end{split}
\end{equation}
Thus,
\begin{equation}
    \begin{split}
        \chi_{\alpha\alpha}^{sp}(J'_{\alpha\alpha}-(2N_{s}-1)^{1/2}\sqrt{M_{\alpha\alpha}^{2}+\sum_{\alpha>\beta}(|J'_{\alpha\beta}|^{2}+|J_{\alpha\beta}|^{2})})	 &\\
        \leq\lambda_{\text{min}}^{\hat{\mathcal{U}}^{sp}\hat{\chi}^{sp}}\leq&\\
\chi_{\alpha\alpha}^{sp}(J'_{\alpha\alpha}-(2N_{s}-1)^{-1/2}\sqrt{M_{\alpha\alpha}^{2}+\sum_{\alpha>\beta}(|J'_{\alpha\beta}|^{2}+|J_{\alpha\beta}|^{2})}),	&
    \end{split}
\end{equation}
meaning the minimum eigenvalue is favored (the lower bound becomes negative) as the number of sublattice sites increases. 
Lastly, the homogeneous orbital elements $\hat{\mathcal{U}}_{ss}\hat{\chi}_{ss}$ are such that $n_{\hat{\mathcal{U}}_{ss}\hat{\chi}_{ss}}=N_{o}N_{s}=N_{b}$, $\text{tr}\hat{\mathcal{U}}_{ss}\hat{\chi}_{ss}=\chi_{\alpha\alpha}^{ss}U_{\alpha\alpha}N_{b}$, and $\text{tr}(\hat{\mathcal{U}}_{ss}\hat{\chi}_{ss})^{2}=(\chi_{\alpha\alpha}^{ss})^{2}(U_{\alpha\alpha}^{2}+N_{\alpha\alpha}^{2}(N_{o}-1)+\sum_{\alpha>\beta}(|U_{\alpha\beta}|^{2}+|U'_{\alpha\beta}|^{2}(N_o-1)))N_{b}$, resulting in parameters
\begin{equation}
    \begin{split}
        m_{\hat{\mathcal{U}}_{ss}\hat{\chi}_{ss}}	=\chi_{\alpha\alpha}^{ss}U_{\alpha\alpha}N_{b}/N_{b}=\chi_{\alpha\alpha}^{ss}U_{\alpha\alpha}, &\\
        s_{\hat{\mathcal{U}}_{ss}\hat{\chi}_{ss}}^{2}	=(\chi_{\alpha\alpha}^{ss})^{2}(U_{\alpha\alpha}^{2}+N_{\alpha\alpha}^{2}(N_{o}-1)&\\
        +\sum_{\alpha>\beta}(|U_{\alpha\beta}|^{2}+|U'_{\alpha\beta}|^{2}(N_o-1)))N_{b}/N_{b}-m_{\hat{\mathcal{U}}^{sp}\hat{\chi}^{sp}}^{2} &\\
	  =(\chi_{\alpha\alpha}^{ss})^{2}(N_{\alpha\alpha}^{2}(N_{o}-1)+\sum_{\alpha>\beta}(|U_{\alpha\beta}|^{2}+|U'_{\alpha\beta}|^{2}(N_o-1)))&.
    \end{split}
\end{equation}
Thus,
\begin{equation}
    \begin{split}
        \chi_{\alpha\alpha}^{ss}\Big\{U_{\alpha\alpha}-(N_{b}-1)^{1/2}&\\
        \big[{N_{\alpha\alpha}^{2}(N_{o}-1)+\sum_{\alpha>\beta}(|U_{\alpha\beta}|^{2}+|U'_{\alpha\beta}|^{2}(N_o-1))}\big]^{1/2}\Big\} &\\
        \leq\lambda_{\text{min}}^{\hat{\mathcal{U}}_{ss}\hat{\chi}_{ss}}
        \leq\chi_{\alpha\alpha}^{ss}\Big\{U_{\alpha\alpha}-(N_{b}-1)^{-1/2}\big[N_{\alpha\alpha}^{2}(N_{o}-1)&\\
        +\sum_{\alpha>\beta}(|U_{\alpha\beta}|^{2}+|U'_{\alpha\beta}|^{2}(N_o-1))\big]^{1/2}\Big\{.&
    \end{split}
\end{equation}
Now, the homogeneous orbital eigenvalues grow with the number of bands and with the Hubbard-like terms, which are typically larger than the exchange terms, suggesting $\max\alpha_{c}=\max-\lambda_{\text{min}}^{\hat{\mathcal{U}}_{ss}\hat{\chi}_{ss}}>\max-\lambda_{\text{min}}^{\hat{\mathcal{U}}_{sp}\hat{\chi}_{sp}}$. 
However, to become negative, the higher bound of $\lambda_{\text{min}}^{\hat{\mathcal{U}}_{sp}\hat{\chi}_{sp}}$ needs to overpass a typically small quantity, $J'_{\alpha\alpha}$. 
Reaching the necessary negative sign, is achieved by two mechanisms. One of them is the contribution of the different-sublattice terms and the other is by increasing the number of orbitals. 
Thus, the sublattice structure or the multiorbital character of a system can cause charge fluctuations to play a role for the system's ground state.

Finally, we compare the influence of charge and spin fluctuations by means of the Stoner parameters derived above. 
In sum, we obtained (changing $\lambda_{\text{min}}^{\hat{\mathcal{U}}_{ss/sp}\hat{\chi}_{ss/sp}}$ by $\lambda_{\text{max}}^{-\hat{\mathcal{U}}_{ss/sp}\hat{\chi}_{ss/sp}}$)
\begin{equation}
\begin{split}
    \lambda_{\text{max}}^{\hat{\mathcal{V}}_{\alpha\alpha}^{sp}\hat{\chi}_{\alpha\alpha}^{sp}}=\chi_{\alpha\alpha}^{sp}(J'_{\alpha\alpha}+U'_{\alpha\alpha})&,
\end{split}
\end{equation}
\begin{equation}
\begin{split}
    \chi_{\alpha\alpha}^{ss}(U_{\alpha\alpha}+J_{\alpha\alpha})\leq\lambda_{\text{max}}^{\hat{\mathcal{V}}_{\alpha\alpha}^{ss}\hat{\chi}_{\alpha\alpha}^{ss}}
    \leq\chi_{\alpha\alpha}^{ss}(U_{\alpha\alpha}+J_{\alpha\alpha}(N_{o}-1)),&
\end{split}
\end{equation}
\begin{equation}
\begin{split}
    \chi_{\alpha\alpha}^{sp}\Big\{-J'_{\alpha\alpha}+(2N_{s}-1)^{-1/2}\big[(-U'_{\alpha\alpha}+2J_{\alpha\alpha})^{2}&\\
    +\sum_{\alpha>\beta}(|J'_{\alpha\beta}|^{2}+|J_{\alpha\beta}|^{2})\big]^{1/2}\Big\}&\\
    \leq\lambda_{\text{max}}^{-\hat{\mathcal{U}}_{sp}\hat{\chi}_{sp}}
    \leq\chi_{\alpha\alpha}^{sp}\Big\{-J'_{\alpha\alpha}+(2N_{s}-1)^{1/2}\big[(-U'_{\alpha\alpha}&\\
    +2J_{\alpha\alpha})^{2}+\sum_{\alpha>\beta}(|J'_{\alpha\beta}|^{2}+|J_{\alpha\beta}|^{2})\big]^{1/2}\Big\},&
\end{split}
\end{equation}
\begin{equation}
\begin{split}
    \chi_{\alpha\alpha}^{ss}\Big\{-U_{\alpha\alpha}+(N_{b}-1)^{-1/2}\big[(2U'_{\alpha\alpha}-J_{\alpha\alpha})^{2}(N_{o}-1)&\\
    +\sum_{\alpha>\beta}(|U_{\alpha\beta}|^{2}+|U'_{\alpha\beta}|^{2}(N_o-1))\big]^{1/2}\Big\}&\\
    \leq\lambda_{\text{max}}^{-\hat{\mathcal{U}}_{ss}\hat{\chi}_{ss}}
    \leq\chi_{\alpha\alpha}^{ss}\Big\{-U_{\alpha\alpha}+(N_{b}-1)^{1/2}\big[(2U'_{\alpha\alpha}&\\
    -J_{\alpha\alpha})^{2}(N_{o}-1)+\sum_{\alpha>\beta}(|U_{\alpha\beta}|^{2}+|U'_{\alpha\beta}|^{2}(N_o-1))\big]^{1/2}\Big\}.&
\end{split}
\end{equation}
Increasing the number of orbitals highers the lower bound of $\max\alpha_{c}$ but does not influence the lower bound of $\max\alpha_{s}$. 
Changing the number of sublattice sites does not influence the boundaries of the spin channel, while it highly favors the charge channel. 
Also, once $J'_{\alpha\alpha}$ is a small quantity, it becomes more likely for the current orbital terms $\lambda_{\text{max}}^{-\hat{\mathcal{U}}_{sp}\hat{\chi}_{sp}}$ to dominate the fluctuations scenario. 
This homogeneous versus current competition clearly shows the strong dependence on the bare susceptibility to the physical state observed in the system, which emphasizes the role of doping for these ground state transitions. 
Next, we analyze the spin-charge fluctuation transitions case by case.
\begin{outline}
    \1 $\max\alpha_{c}>\max\alpha_{s}$ case (lower bound of $\max\alpha_{c}>$ higher bound of $\max\alpha_{s}$)
        \2 If $\chi_{\alpha\alpha}^{ss}$ is the main channel, then $\lambda_{\text{max}}^{-\hat{\mathcal{U}}_{ss}\hat{\chi}_{ss}}$ and $\lambda_{\text{max}}^{\hat{\mathcal{V}}_{\alpha\alpha}^{ss}\hat{\chi}_{\alpha\alpha}^{ss}}$ compete:
\end{outline}
        \begin{equation}
            \begin{split}
                \chi_{\alpha\alpha}^{ss}\Big\{-U_{\alpha\alpha}+(N_{b}-1)^{-1/2}\big[(2U'_{\alpha\alpha}-J_{\alpha\alpha})^{2}(N_{o}-1)&\\
                +\sum_{\alpha>\beta}(|U_{\alpha\beta}|^{2}+|U'_{\alpha\beta}|^{2}(N_o-1))\big]^{1/2}\Big\} & \\
                >\chi_{\alpha\alpha}^{ss}\big[U_{\alpha\alpha}+J_{\alpha\alpha}(N_{o}-1)\big] &
            \end{split}
        \end{equation}
        \begin{equation}
            \begin{split}
            (2U'_{\alpha\alpha}-J_{\alpha\alpha})^{2}\frac{(N_{o}-1)}{(N_{b}-1)}
            +\frac{1}{(N_{b}-1)}\sum_{\alpha>\beta}(|U_{\alpha\beta}|^{2}&\\
            +|U'_{\alpha\beta}|^{2}(N_o-1))>(2U_{\alpha\alpha}+J_{\alpha\alpha}(N_{o}-1))^{2}.&
            \end{split}
        \end{equation}
\begin{outline}
    \1\3 The increasing number of different sublattice sites plus the lattice geometry can make the$ \sum_{\alpha>\beta}(|U_{\alpha\beta}|^{2}+|U'_{\alpha\beta}|^{2})$ term relevant enough to induce the spin-to-charge transition. 
    \1 If $N_{s}=1$, then the transition never happens. 
    Sublattice geometry is, therefore, essential for the dominating charge fluctuations.
    \2 If $\chi_{\alpha\alpha}^{sp}$ is the main channel, then $\lambda_{\text{max}}^{-\hat{\mathcal{U}}_{sp}\hat{\chi}_{sp}}$ and $\lambda_{\text{max}}^{\hat{\mathcal{V}}_{\alpha\alpha}^{sp}\hat{\chi}_{\alpha\alpha}^{sp}}$ compete:
\end{outline}
\begin{equation}
    \begin{split}
        \chi_{\alpha\alpha}^{sp}\Big\{-J'_{\alpha\alpha}+(2N_{s}-1)^{-1/2}\big[(-U'_{\alpha\alpha}+2J_{\alpha\alpha})^{2}&\\
        +\sum_{\alpha>\beta}(|J'_{\alpha\beta}|^{2}+|J_{\alpha\beta}|^{2})\big]^{1/2}\Big\}>\chi_{\alpha\alpha}^{sp}(J'_{\alpha\alpha}+U'_{\alpha\alpha})&
    \end{split}
\end{equation}
\begin{equation}
    \begin{split}
        (-U'_{\alpha\alpha}+2J_{\alpha\alpha})^{2}+\sum_{\alpha>\beta}(|J'_{\alpha\beta}|^{2}+|J_{\alpha\beta}|^{2})>&\\
        (2N_{s}-1)(2J'_{\alpha\alpha}+U'_{\alpha\alpha})^{2}.&
    \end{split}
\end{equation}
\begin{outline}
            \1\3 As in the dominating homogeneous channel case, there is no transition without non-trivial sublattice geometry.
\end{outline}
\begin{outline}
    \1 $\max\alpha_{s}>\max\alpha_{c}$ case (lower bound of $\max\alpha_{s}>$ higher bound of $\max\alpha_{c}$):
        \2 If $\chi_{\alpha\alpha}^{ss}$ is the main channel, then $\lambda_{\text{max}}^{-\hat{\mathcal{U}}_{ss}\hat{\chi}_{ss}}$ and $\lambda_{\text{max}}^{\hat{\mathcal{V}}_{\alpha\alpha}^{ss}\hat{\chi}_{\alpha\alpha}^{ss}}$ compete:
\end{outline}
\begin{equation}
    \begin{split}
        \chi_{\alpha\alpha}^{ss}\Big\{U_{\alpha\alpha}+J_{\alpha\alpha})>\chi_{\alpha\alpha}^{ss}(-U_{\alpha\alpha}+&\\
        (N_{b}-1)^{1/2}\big[(2U'_{\alpha\alpha}-J_{\alpha\alpha})^{2}(N_{o}-1)&\\
        +\sum_{\alpha>\beta}(|U_{\alpha\beta}|^{2}+|U'_{\alpha\beta}|^{2}(N_o-1))\big]^{1/2}\Big\}
    \end{split}
\end{equation}
\begin{equation}
    \begin{split}
        (2U_{\alpha\alpha}+J_{\alpha\alpha})^{2}>(2U'_{\alpha\alpha}-J_{\alpha\alpha})^{2}(N_{b}-1)(N_{o}-1)&\\
        +(N_{b}-1)\sum_{\alpha>\beta}(|U_{\alpha\beta}|^{2}+|U'_{\alpha\beta}|^{2}(N_o-1)).&
    \end{split}
\end{equation}
\begin{outline}
            \1\3 The spin fluctuations dominate when the number of bands or orbitals are low (multiband character disadvantages dominating spin fluctuations) or if $J_{\alpha\alpha}$ is large. 
            Also, sublattice geometry disadvantages the spin channel, as previously concluded.
        \1 If $\chi_{\alpha\alpha}^{sp}$ is the main channel, then $\lambda_{\text{max}}^{-\hat{\mathcal{U}}_{sp}\hat{\chi}_{sp}}$ and $\lambda_{\text{max}}^{\hat{\mathcal{V}}_{\alpha\alpha}^{sp}\hat{\chi}_{\alpha\alpha}^{sp}}$ compete:
\end{outline}
\begin{equation}
    \begin{split}
        \chi_{\alpha\alpha}^{sp}(J'_{\alpha\alpha}+U'_{\alpha\alpha})>\chi_{\alpha\alpha}^{sp}\Big\{-J'_{\alpha\alpha}&\\
        +(2N_{s}-1)^{1/2}\big[(-U'_{\alpha\alpha}+2J_{\alpha\alpha})^{2}&\\
        +\sum_{\alpha>\beta}(|J'_{\alpha\beta}|^{2}+|J_{\alpha\beta}|^{2})\big]^{1/2}\Big\}&
    \end{split}
\end{equation}
\begin{equation}
    \begin{split}
        (2J'_{\alpha\alpha}+U'_{\alpha\alpha})^{2}>(2N_{s}-1)(-U'_{\alpha\alpha}+2J_{\alpha\alpha})^{2}&\\
        +(2N_{s}-1)\sum_{\alpha>\beta}(|J'_{\alpha\beta}|^{2}+|J_{\alpha\beta}|^{2}).&
    \end{split}
\end{equation}
\begin{outline}
            \1\3 Again, the dominance of spin fluctuations in the current channels is fragile to the number of sublattice sites.
\end{outline}
figure~\ref{fig:fluctuations} in the main text shows a schematics of possible spin-charge interplay scenarios.

Performing a similar analysis for the kernel $\tilde{\Gamma}(\boldsymbol{k},\boldsymbol{k}')=-\frac{d\boldsymbol{k}'_{||}}{v_{F}(\boldsymbol{k}')}\frac{1}{(2\pi)^{2}}\Gamma(\boldsymbol{k},\boldsymbol{k}')$, which has as eigenvalues the real-valued pairing strengths $\lambda_{\boldsymbol{k}}$ (such that the bound theorem holds), we obtain by considering equal number $N_{\boldsymbol{k}}$ of $\boldsymbol{k}$ and $\boldsymbol{k}'$ points,
\begin{equation}
    \begin{split}
        m_{\tilde{\Gamma}(\boldsymbol{k},\boldsymbol{k}')}=\text{tr}\tilde{\Gamma}(\boldsymbol{k},\boldsymbol{k}')/n_{\tilde{\Gamma}(\boldsymbol{k},\boldsymbol{k}')}=\frac{1}{N_{\boldsymbol{k}}}\sum_{\boldsymbol{k}}\tilde{\Gamma}(\boldsymbol{k},\boldsymbol{k}),&\\
        s_{\tilde{\Gamma}(\boldsymbol{k},\boldsymbol{k}')}^{2}=\text{tr}\tilde{\Gamma}(\boldsymbol{k},\boldsymbol{k}')^{2}/n_{\tilde{\Gamma}(\boldsymbol{k},\boldsymbol{k}')}-m_{\tilde{\Gamma}(\boldsymbol{k},\boldsymbol{k}')}^{2}&\\
        =\frac{1}{N_{\boldsymbol{k}}}\sum_{\boldsymbol{k},\boldsymbol{k}'}\tilde{\Gamma}(\boldsymbol{k},\boldsymbol{k}')\tilde{\Gamma}(\boldsymbol{k}',\boldsymbol{k})&\\
        -\frac{1}{N_{\boldsymbol{k}}}\left(\sum_{\boldsymbol{k}}\tilde{\Gamma}(\boldsymbol{k},\boldsymbol{k})\right)^{2}.
    \end{split}
\end{equation}
Thus, the lower bound for the maximum pairing strength is given by
\begin{equation}
    \begin{split}
    \frac{1}{N_{\boldsymbol{k}}}\sum_{\boldsymbol{k}}\tilde{\Gamma}(\boldsymbol{k},\boldsymbol{k})&\\
    +\Bigg[\frac{1}{N_{\boldsymbol{k}}(N_{\boldsymbol{k}}-1)}\sum_{\boldsymbol{k},\boldsymbol{k}'}\tilde{\Gamma}(\boldsymbol{k},\boldsymbol{k}')\tilde{\Gamma}(\boldsymbol{k}',\boldsymbol{k})&\\
    -\frac{1}{N_{\boldsymbol{k}}^{2}(N_{\boldsymbol{k}}-1)}\left(\sum_{\boldsymbol{k}}\tilde{\Gamma}(\boldsymbol{k},\boldsymbol{k})\right)^{2}\Bigg]^{1/2}	\leq\lambda_{\text{max}}^{\tilde{\Gamma}(\boldsymbol{k},\boldsymbol{k}')}=\lambda.&\\
    \end{split}
\end{equation}
writing as a function of $\Gamma(\boldsymbol{k},\boldsymbol{k}')$ we obtain
\begin{equation}
    \begin{split}
        \lambda\geq-\sum_{\boldsymbol{k}}\frac{d\boldsymbol{k}_{||}}{v_{F}(\boldsymbol{k})}\frac{1}{N_{\boldsymbol{k}}(2\pi)^{2}}\Gamma(\boldsymbol{k},\boldsymbol{k})&\\
        +\Bigg[\frac{1}{N_{\boldsymbol{k}}(N_{\boldsymbol{k}}-1)}\sum_{\boldsymbol{k},\boldsymbol{k}'}\frac{d\boldsymbol{k}'_{||}d\boldsymbol{k}_{||}}{v_{F}(\boldsymbol{k}')v_{F}(\boldsymbol{k})}\frac{1}{(2\pi)^{4}}\Gamma(\boldsymbol{k},\boldsymbol{k}')\Gamma(\boldsymbol{k}',\boldsymbol{k})&\\
        -\frac{1}{N_{\boldsymbol{k}}^{2}(N_{\boldsymbol{k}}-1)}\left(\sum_{\boldsymbol{k}}\frac{d\boldsymbol{k}{}_{||}}{v_{F}(\boldsymbol{k})}\frac{1}{(2\pi)^{2}}\Gamma(\boldsymbol{k},\boldsymbol{k})\right)^{2}\Bigg]^{1/2}.&
    \end{split}
\end{equation}
Therefore, the lower bound for $\lambda$ is enhanced by $\boldsymbol{q}\neq 0$ fluctuations. 
If several channels are presenting $\boldsymbol{q}\neq 0$ peaks, then $\lambda$ is expected to increase, favoring stronger coupling superconductivity.

\end{appendix}

\bibliographystyle{iopart-num}
\bibliography{tbg,kagome,nickelates,TBG_v2,TBG_v3}

\end{document}